\documentclass[a4paper,fleqn]{cas-sc}

\usepackage[authoryear,longnamesfirst]{natbib}
\usepackage{amsmath}
\usepackage{amsfonts}
\usepackage{amssymb}
\usepackage{caption}
\usepackage{subcaption}
\usepackage{graphicx}
\def\tsc#1{\csdef{#1}{\textsc{\lowercase{#1}}\xspace}}
\tsc{WGM}
\tsc{QE}

\newtheorem{Def}{Definition}[section]

\begin{document}
\let\WriteBookmarks\relax
\def\floatpagepagefraction{1}
\def\textpagefraction{.001}

 \shorttitle{Zigzag and Checkerboard Patterns in Coupled Maps}    

 \shortauthors{Warambhe and Gade}  

\title [mode = title]{Approach to zigzag and checkerboard patterns in 
 spatially extended systems}  



%

\author[1]{Manoj C. Warambhe} 



\ead{manojwarambhe8@gmail.com}


\credit{Conceptualization of this study, Methodology, Software}


\author[1]{Prashant M. Gade}[]
\cormark[1]

\ead{prashant.m.gade@gmail.com}

\ead[url]{}

\credit{Conceptualization of this study, Methodology, Software}

\affiliation[1]{organization={Department of Physics},
            addressline={Rashtrasant Tukadoji Maharaj Nagpur University}, 
            city={Nagpur},
            postcode={440033}, 
            state={Maharashtra},
            country={India}}

\cortext[1]{ca}

\fntext[2]{https://orcid.org/0000-0002-8343-823X}


\begin{abstract}
	Zigzag patterns in one dimension or checkerboard patterns in 
	two dimensions occur in a variety of pattern-forming systems.
	We introduce an order parameter `phase defect'
	to identify this transition and help
	to recognize the associated universality class on a discrete lattice.
	In one dimension, if $x_{i}(t)$ is a variable value at site $i$ at
	time $t$. We assign spin $s_i(t)=1$ for $x_{i}(t)>x_{i-1}(t)$, $s_i(t)=-1$
	if $x_{i}(t)<x_{i-1}(t)$, and $s_i(t)=0$
	if $x_{i}(t)=x_{i-1}(t)$. 
	The phase defect $D(t)$ is defined as $D(t)={\frac{\sum_{i=1}^N \vert
	s_i(t)+s_{i-1}(t)\vert} {2N}}$  for  a
	a lattice of $N$ sites with periodic boundary conditions.
	It is zero for a zigzag pattern. In two dimensions, $D(t)$ is the sum of
	row-wise as well as column-wise phase defects and is zero for the checkerboard pattern. 
	The persistence $P(t)$ is the fraction of sites whose spin value did not change even once till time $t$.
	We find that $D(t)\sim t^{-\delta}$ and $P(t)\sim t^{-\theta}$
	for the parameter range over which
	the zigzag or checkerboard pattern is realized.
	We observe that $\delta=0.5$ and $\theta=3/8$ for 1-d coupled logistic maps
	or Gauss maps and $\theta=0.22$ and $\delta=0.45$ in 2-d logistic or Gauss maps.
	The exponent $\theta$ matches with the persistence exponent at
	zero temperature for the Ising model and 
	$\delta$ matches with the exponent for the Ising model at the critical temperature.
	This power-law decay is observed over a range of parameter values and not just critical point.
\end{abstract}


\begin{highlights}
\item We define an order parameter for zigzag and checkerboard patterns on a discrete lattice.
\item We find that the order parameter shows a power-law decay over a range of parameter values and not just critical point.
\item  We define persistence in these systems and it shows power-law decay in the same range with exponents similar to those of the Ising model.
\end{highlights}


\maketitle

\section{Introduction}
Zigzag and checkerboard patterns are observed in a variety of 
spatially extended systems. 
	Such patterns are observed 
	in molecular dynamics simulations as well as in 
	experiments \cite{mielenz2013trapping}. 
If ions are confined in a beam,
the configuration of cold ions forms a zigzag pattern with increasing 
	density \cite{schiffer1993phase}. 
	Zigzag bifurcation
	is observed in bulk lamellar eutectic growth \cite{akamatsu2004experimental}.
	Zigzag walls are observed in nematic liquid crystals
	with negative dielectric anisotropy \cite{nagaya2002experimental}.
	Zigzag destabilization of self-organized solitary stripes was
	detected in the current density of a planar semiconductor gas discharge system \cite{strumpel2000dynamics}.
	A zigzag pattern of droplets is observed in vesicle-generating
	microfluidic device \cite{thorsen2001dynamic}. 
	One-dimensional cellular neural networks made of Chua's circuits show zigzag patterns \cite{zheleznyak1994coexistence}. 
	This instability is observed in intestine growth in several embryos ranging from humans to fishes \cite{ben2013anisotropic}.
	This is not surprising because zigzag instability is
	one of the basic instability routes in pattern-forming systems \cite{cross1993pattern}. 
	Checkerboard patterns are observed in the distribution of species pairs due to 
	competitive exclusion \cite{stone1990checkerboard}. They are observed in chemical 
	reactions \cite{horvath2009experimental} and even in
	coupled genetic oscillators on a chip of artificial cells \cite{tayar2017synchrony}.
	Thus these patterns occur widely in models as well as experiments. 
	
	We study these patterns in a much simpler system of coupled chaotic maps.
	Coupled logistic maps in one and two dimensions were studied extensively by Kaneko who can be credited for making this
	field popular. In one dimension, a zigzag pattern is observed
	over a large parameter region \cite{kaneko1988chaotic}. Similarly,
	a checkerboard pattern is observed in two dimensions over a
	wide range \cite{kaneko1989spatiotemporal}.
	Often the reference is for the fixed point\cite{gade2013} and
	the comparison is between two successive time steps. Another
	definition considers two successive time-steps 
	for single as well as coupled map is given by \cite{bambi}. 
	We are interested in spatial order and give definitions by considering
	the local spatial slope for each site in this work.

	Diverse spatiotemporal phases are observed in CML and transitions
between those have been studied as dynamic phase transitions.
One of the most studied transitions is the transition to a synchronized state.
Directed percolation (DP) transition has been observed in the transition from
spatiotemporal intermittency to laminar state in one-dimensional \cite{chatterjee1996synchronization} as well as two-dimensional
coupled map \cite{chate1988continuous}. Even for states lacking long-range spatial order DP class is obtained
in transition to a coarse-grained frozen state in coupled Gauss maps \cite{pakhare2020novel}.
Replica synchronization of coupled maps is in multiplicative noise universality class in 
one dimension \cite{ahlers2002critical} as well as in
two dimensions \cite{ginelli2009synchronization}. 
If there are multiple absorbing states linked by certain symmetry, other classes can be obtained. For some maps, the transition is
        observed in $q=3$ potts universality class \cite{salazar2005critical}.
        For maps with Ising symmetry, Miller and Huse showed that the transition is in 
	the Ising universality class \cite{miller1993macroscopic}.
        Map with long delay can be mapped on pseudo-spatiotemporal
	system {\it{i.e.}} CML with a typewriter mode
	update in typewriter mode. Directed
        Ising  (DI) transition is observed in the logistic map with
        delay \cite{mahajan2019transition}. For the $\mathcal{Z}_2$ symmetry, Ising-type or DI  transition can be obtained.
	In the former case, the order parameter decays with the exponent
        $1/2$ at the critical point and latter case the exponent is
        $0.285$\cite{nitesh,deshmukh2021effect}.
        The transitions are also studied using persistence as an order parameter. 
	For coupled logistic maps with delay and linear
        or nonlinear couplings, ten different systems have been studied
        which fall into two different universality classes depending
        on if the nature of long-range order obtained at the transition
        is ferromagnetic or antiferromagetic\cite{rajvaidya2020transition}.
	
	In all these cases, there is a critical point and nontrivial
power-laws are obtained only at a critical point.
Both in simulations as well as in experiments, it is an arduous and
time-consuming task to fine-tune the control parameter
and the power law. However, we know that
power laws in space and time are abundant in nature\cite{bak2013nature}.
Thus nature finds it effortlessly. A couple of mechanisms are proposed to explain 
the observation of such power laws without specific tuning of parameters.
One is self-organized criticality\cite{bak2013nature}.
It requires a slow adiabatic drive.
Another mechanism is Griffiths phase which deals with
quenched disorder. It is present in several systems and this
explanation certainly has physical relevance.
Such a system can show continuously varying power laws in a range of parameter values\cite{vojta2006rare,bhoyar2020dynamic}
and the exponents can even be complex.

The striking result in our study on zigzag and checkerboard patterns
is that both persistence as well as phase defects decay as a power law over a large parameter range and not just a critical point. We proceed as follows. We find the local slope and
take zero as a reference. Thus the sites which have a value larger(smaller)
than left hand side neighbor have +1(-1) spin value.
A zigzag pattern will have a strictly alternating sequence of +1 and -1
spins. We consider deviation from it as a phase defect and consider
it as an order parameter. The same definition is extended to two dimensions. Similarly, we introduce phase persistence. All sites which have not changed
the spin value from the initial state all even times till time $t$
are persistent sites. Visualization and analysis of zigzag patterns are usually with respect to a fixed point. This definition needs the knowledge of underlying equations. Besides, such a quantifier underestimates the range over which zigzag patterns are actually observed and hence we introduce this order parameter.

The studies from a viewpoint of phase transition are identified by a suitable order parameter. Several dynamical phases are possible in spatiotemporal systems and they can be identified from spatiotemporal patterns. However, visual inspection over the entire space is cumbersome and certain scalar or vector quantifiers can help us to identify these phases. Except for trivial phases such as synchronized state or periodic
state, it is not easy to identify these quantifiers.

\begin{Def}
For the zigzag pattern, we define
phase defect in space in the 1-d lattice in the following manner.
	Let $x_{i}(t)$ be the variable value at site $i$ at time $t$.
	We assign $s_i(t)=1$ for $x_{i}(t)>x_{i-1}(t)$,  $s_i(t)=-1$
	if  $x_{i}(t)<x_{i-1}(t)$, and $s_i(t)=0$
	if  $x_{i}(t)=x_{i-1}(t)$.
The phase defect $D(t)$ is the fraction of two adjacent
sites associated with the same spin given by

\begin{equation*} 
	D(t)={\frac{1}{2N}}\sum_{i=1}^N \vert s_{i}(t)+s_{i-1}(t)\vert
\end{equation*}
\end{Def}

\begin{Def}
The phase persistence $P(t)$ is the
number of sites which did not change their spin even once and for all even time steps. It is fraction of sites for which $s_i(t')=s_i(0)$ at all
even times $t' \le t$.
\end{Def}

\begin{Def}
In two dimensions, let $x(i,j)$ be the variable value at sites $i$ and $j$.
If $x_{i,j}(t) > x_{i-1,j}(t)$, we associate spin $s_{i,j}(t)=1$ with this site, and if $x_{i,j}(t)<x_{i-1,j}(t)$ we associate spin $s_{i,j}(t)=-1$ with it and
if $x_{i,j}(t)=x_{i-1,j}(t)$ we associate spin $s_{i,j}(t)=0$.
Due to the two-dimensional nature of the map, we associate
one more spin $s'_{i,j}$ with site $(i,j)$. If $x_{i,j}(t) > x_{i,j-1}(t)$, we assign spin $s'_{i,j}(t)=1$ with this site, and if $x_{i,j}(t)<x_{i,j-1}(t)$
we assign  spin $s'_{i,j}(t)=-1$ with it and if $x_{i,j}(t)=x_{i,j-1}(t)$
we assign  spin $s'_{i,j}(t)=0$. Now the phase defect is a sum of phase defects  in both variables $s$ and $s'$ and is defined as

\begin{equation*}
	D(t)=
{\frac{1}{2N^2}}
	(\sum_{i,j=1}^N 
	\vert s_{i,j}(t)+s_{i-1,j}(t)\vert
	+\sum_{i,j=1}^N \vert s'_{i,j}(t)+s'_{i,j-1}(t)\vert)
\end{equation*}
\end{Def}

\begin{Def}
We also compute phase persistence in a similar manner. If
$s_{i,j}$ has not changed sign even once for all even times till time $t$,
it has persisted till time $t$. Similarly if $s'_{i,j}$ has not changed sign even once for all even times till time $t$,
it has persisted till time $t$. Phase persistence is the sum of both types of spins that have persisted till time $t$.
\end{Def}

A note on studies on persistence in statistical physics, in general, \cite{majumdar} and in coupled map lattices, in particular, may be in order.
 This is a non-Markovian quantity. In several systems, the persistence probability decays algebraically with time as $P(t)\sim {1}/{t^{\theta}}$
at the critical point where $\theta$ is the decay exponent
known as the persistence exponent. In spin systems such as Ising or Potts spins, the persistence is defined as the probability that a given spin has not flipped from its initial state even once up to
time $t$ \cite{menon2003persistence}. In these systems,
the persistence shows power-law behaviour only at zero temperature.
In one dimension the exponent is $3/8$
 \cite{derrida1994non,derrida1995exact}
and is $0.22$ in two dimensions\cite{stauffer1994ising}.
The persistence exponent of different models such as the 1-d Ising model, coupled logistic maps and the Sznajd model is
found to be
$3/8$\cite{majumdar,stauffer2002persistence,gade2013universal}.
The persistence exponent is not considered a universal exponent.
Still, the persistent exponent of the 1-d DP model such as the Domany-Kinzel model
\cite{hinrichsen1998numerical}, Ziff-Gulari-Barshed model
\cite{albano2001numerical}, site percolation \cite{fuchs2008local},
coupled circle maps and coupled Gauss maps in
one dimension is 3/2 \cite{menon2003persistence,pakhare2020novel}.
Similarly, several models in the directed Ising universality class show
persistence exponent one\cite{shambha}.
For coupled maps with delay, ten different systems have been
studied and the persistence exponent is either
$3/8$ or $2/7$ depending on if the state has long-range
antiferromagnetic or ferromagnetic ordering in the coarse-grained sense.
However, these power laws are seen only at a critical point.
Only for coupled circle maps with repulsive coupling
a persistence exponent $1$ is obtained over a range of
parameters where the travelling wave state is obtained\cite{gade2007power}.
In this work, we obtain power laws in both order parameter and
persistence over a range of parameters.

\section{The model}
\begin{figure*}[h]
       \includegraphics[scale=0.3]{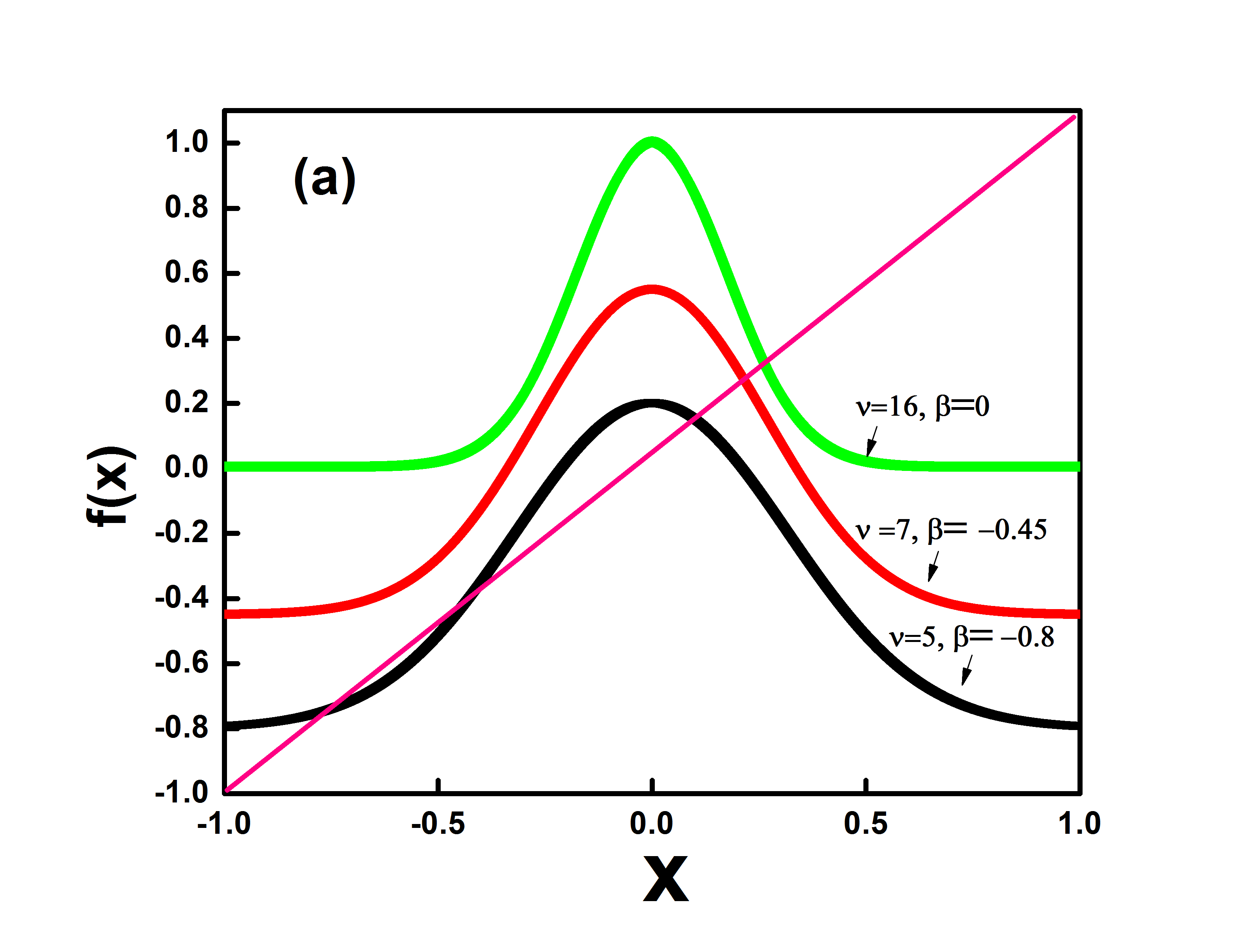}
       \includegraphics[scale=0.3]{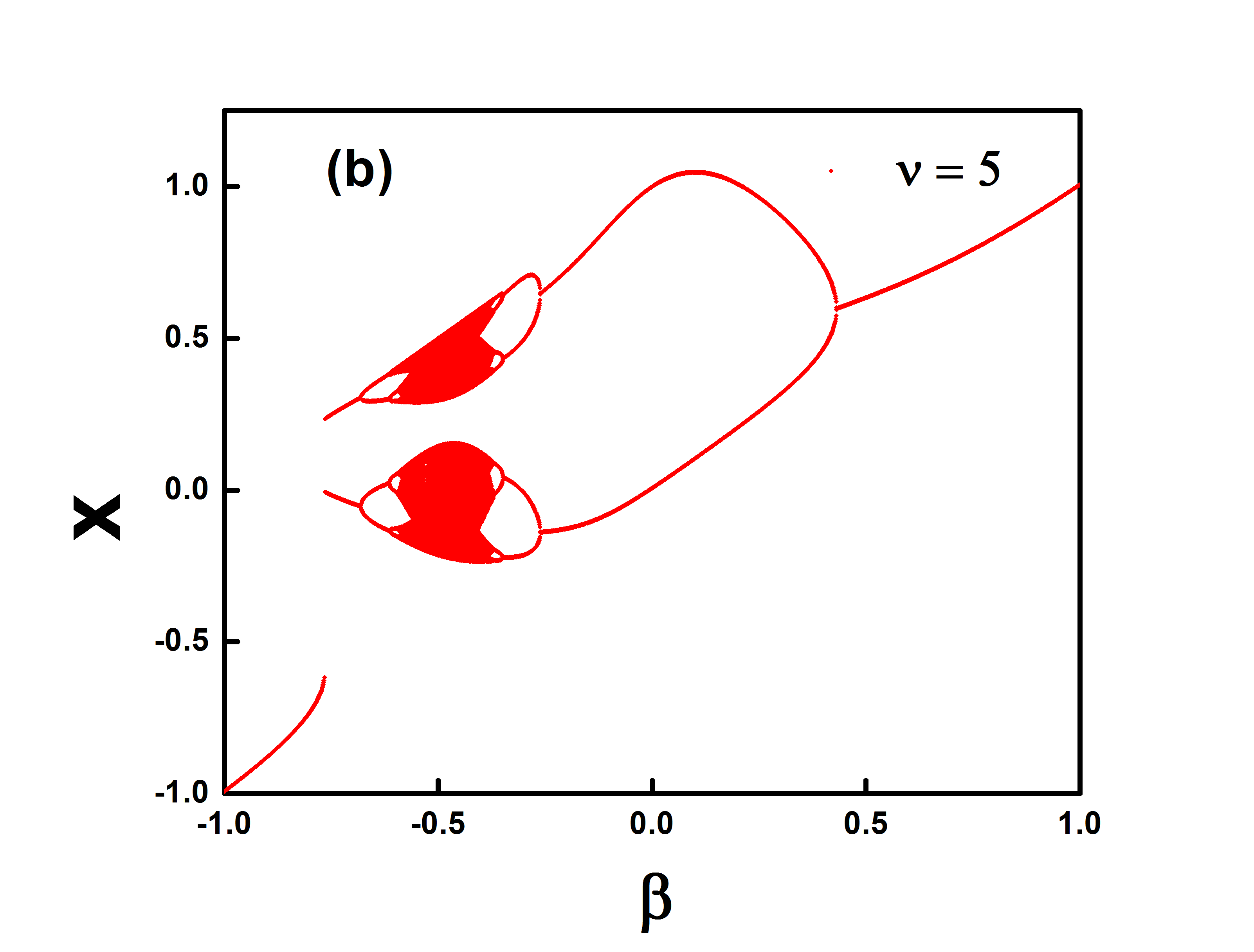}
       \includegraphics[scale=0.3]{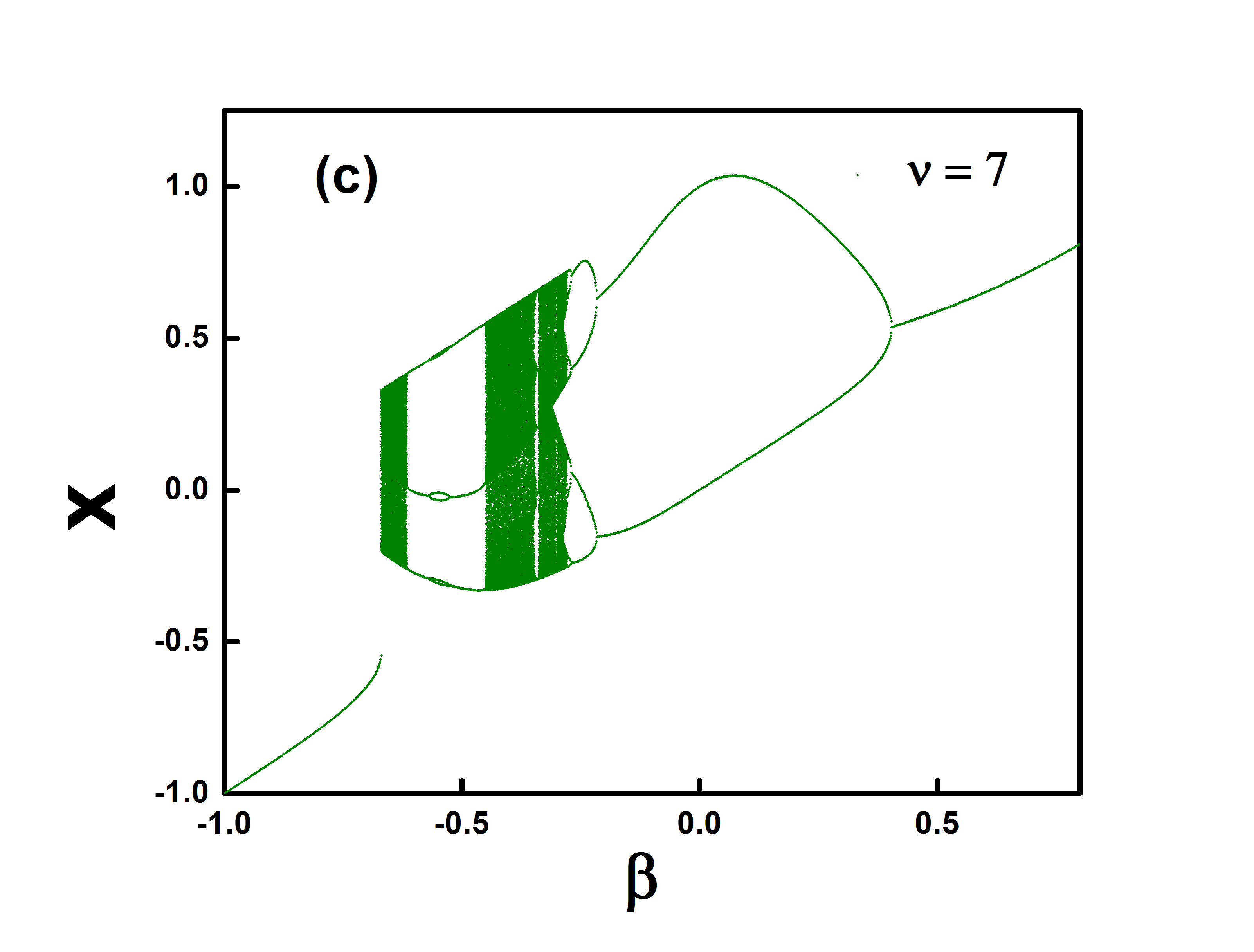}
       \includegraphics[scale=0.3]{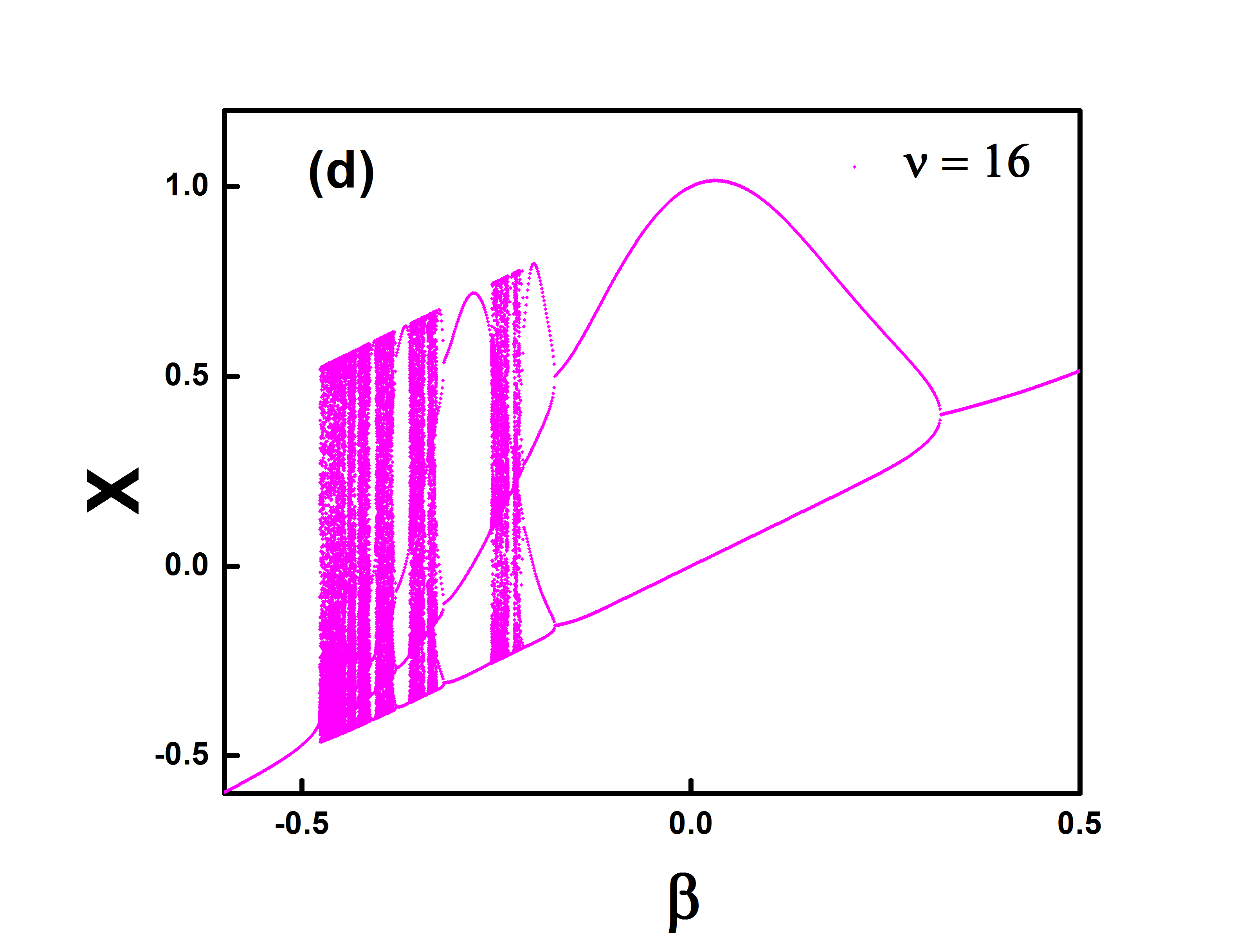}
       \caption{(a) Gauss map is plotted for
$\nu=5, \beta=-0.8$, $\nu=7,\beta=-0.45$ and $\nu=16,\beta=0$.
        For smaller $\beta$, we observe 3 fixed points while for
        larger $\beta$, we observe only one.
       b) Bifurcation diagram for single Gauss map
        as function of $\beta$ is plotted for $\nu=5$.
        We observe reverse period-doubling. c) Same plot for
        $\nu=7$ d) Bifurcation diagram for  $\nu=16$. We observe
        period adding, interspersed by the chaotic attractor.}
       \label{fig1}
\end{figure*}

We consider coupled map lattice with two representative maps.
One is a logistic map which is a well-known map on an interval.
Another is the Gauss map. The Gauss map is a one-dimensional map based on the Gaussian exponential function\cite{hilborn2000chaos}.
On changing parameter values, this map shows a range of dynamical
behaviours including period doubling, reverse period doubling, period adding as well as chaos. Thus new bifurcations that are not seen in maps on an interval are obtained.
It is defined by the following equation.
\begin{equation}
f(x)=\exp(-\nu x^2)+\beta, \hspace{1cm} x\in R
\label{eq1}
\end{equation}
Where $\nu$ and $\beta$ are the map parameters.
The value of function $f(x)$ tends to $\beta$ as
$x\rightarrow\infty$. As we decrease $\beta$, the number of fixed points
changes from 1 to 3. In Fig. \ref{fig1} a), we have shown a Gauss map for various values of $\nu$ and $\beta$.
We have shown the bifurcation diagram for
three different values of $\nu$ in Fig. \ref{fig1} b),c),d).
We observe inverse period doubling and period adding interspersed
by chaotic attractors in these bifurcation diagrams. 
Thus the bifurcation diagram of the Gauss map shows several features not seen in maps on interval.

Secondly, we study the logistic map. The standard logistic map is defined as
\begin{equation}
f(x)=\mu x (1-x),   \hspace{1cm} x\in R
\label{eq2}
\end{equation}
where $\mu$ is the map parameter and the bifurcation diagram is well known. In our work, we fix $\mu=4$. 

We define the coupled map lattice model in one and two dimensions
for completeness. Consider a lattice of length $N$.
We associate a continuous variable $x_{i}(t)$ at each site $i$ at 
time $t$. The evolution equation is,
\begin{equation}
        x_i(t+1)=(1-\epsilon)f(x_i(t))+{\frac{\epsilon}{2}}
	(f(x_{i+1}(t))+f(x_{i-1}(t)))
\label{eq3}
\end{equation}
where, $1 \le i \le N$. 
Now, consider a lattice of size $N \times N$. 
The discrete-time
evolution of 2-d coupled map lattice is,
\begin{equation}
	x_{i,j}(t+1)=(1-\epsilon)f(x_{i,j}(t))+{\frac{\epsilon}{4}}
        (f(x_{i+1,j}(t))+f(x_{i,j+1}(t))+f(x_{i-1,j}(t))+f(x_{i,j-1}(t)))
\label{eq4}
\end{equation}
Where, $i$ and $ j $ range from 1 to $N$. 
We assume periodic boundary conditions in both cases.
The coupling parameter $\epsilon$ is the 
the measure of the strength of diffusive coupling between  site 
and its neighbours. The coupling may be repulsive and attractive. The maps on an interval do not allow negative coupling. But for the Gauss
map, we can study negative values of $\epsilon$. 

Previously, there have been studies on persistence in spin systems
as well as in coupled map lattice models that are mapped on 
spin systems. In these cases, the persistence saturates in a certain
regime and decays to zero elsewhere. There exists a
critical point separating the nonzero asymptotic value of persistence and the zero asymptotic value of persistence.
When persistence saturates, the order parameter goes to zero and when persistence goes to zero, the order parameter saturates. 
The non-zero persistence may be due to a frozen state indicating that a fraction of sites never turns the spin value from their initial conditions.
As noted above, only at the critical point, both persistence and order parameter show a power-law decay.
This has been observed in several cases both in
spin systems as well as in couple map lattices\cite{majumdar,stauffer2002persistence,gade2013universal}.

The difference between the coarse-graining in previous works and this work is the following. In previous works, the spin value is assigned based on the variable value. The spin value is positive (negative) if the variable
value is above (below) the fixed point. On the other hand, in our formulation, the spin value is assigned based on the
sign of difference of nearest neighbours and without any reference to
any special point. Thus it is not necessary to know the underlying equations of the system.
Apart from that, we observe power law over a range of parameters.

\section{Power-laws over a range}
\subsection{Gauss-map with repulsive coupling}
\begin{figure*}[h]
       \centering
       \includegraphics[scale=0.35]{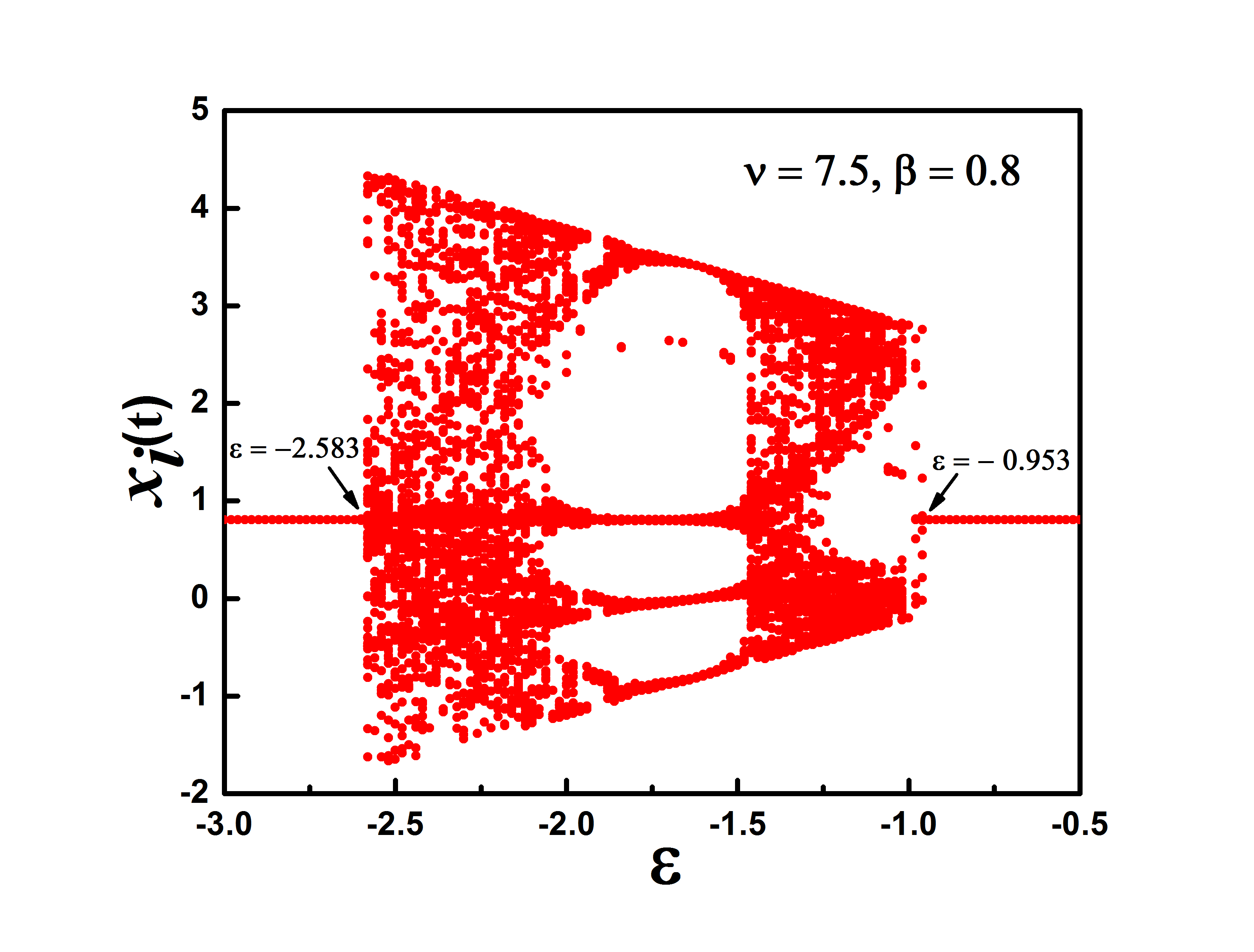}
       \caption{Bifurcation diagram of coupled Gauss map for $\nu=7.5$ and
	$\beta=0.8$ in which all sites $x_{i}(t)$ are plotted as a function 
	repulsive coupling $\epsilon$ at $t=1000$ and $N=100$.}
       \label{fig2}
\end{figure*}

A bifurcation diagram is a simple way in which qualitative changes in
dynamics can be observed at a glance. In this diagram, we plot the values of all sites as a function of the appropriate bifurcation parameter. 
We plot such a diagram for Gauss-map as a function of
repulsive coupling parameter $\epsilon$(see Fig.\ref{fig2}). 
The fixed point becomes unstable at
$\epsilon=-0.953$ and $\epsilon=-2.583$.
Below $\epsilon=-0.953$ we observe 2 bands that merge 
again at $\epsilon=-1.264$. From the bifurcation diagram, we cannot
infer completely about spatiotemporal structure. For example,
at $\epsilon=-1.637$, it is difficult to know if a 4-period attractor 
is coexisting with the fixed point or 5-period state etc. 
We cannot know if there are patterns with certain
spatial order. Thus though
useful, a bifurcation diagram has its limitations and certain
order parameters can be useful in identifying specific phases.

\begin{figure*}[h]
       \includegraphics[scale=0.3]{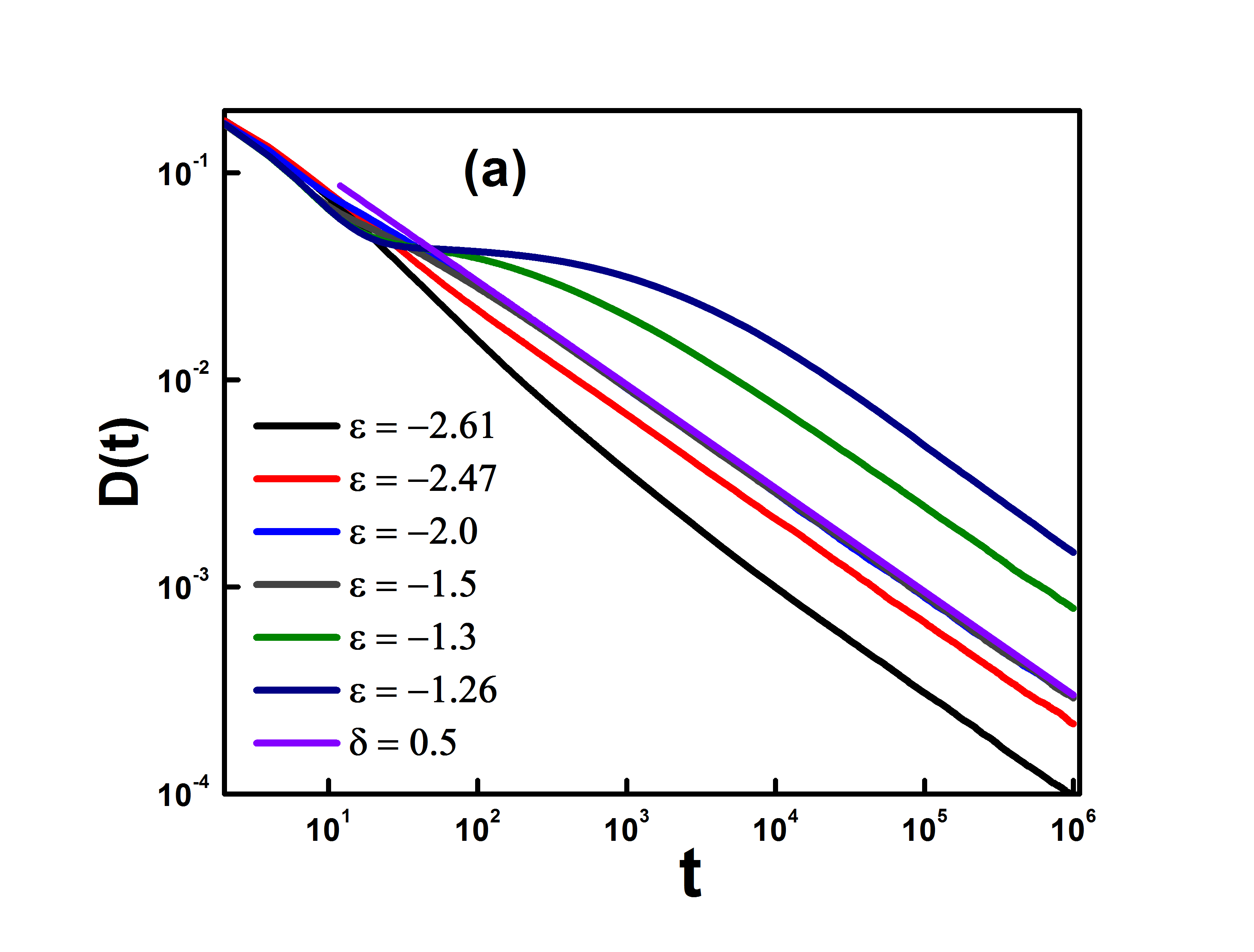}
       \includegraphics[scale=0.3]{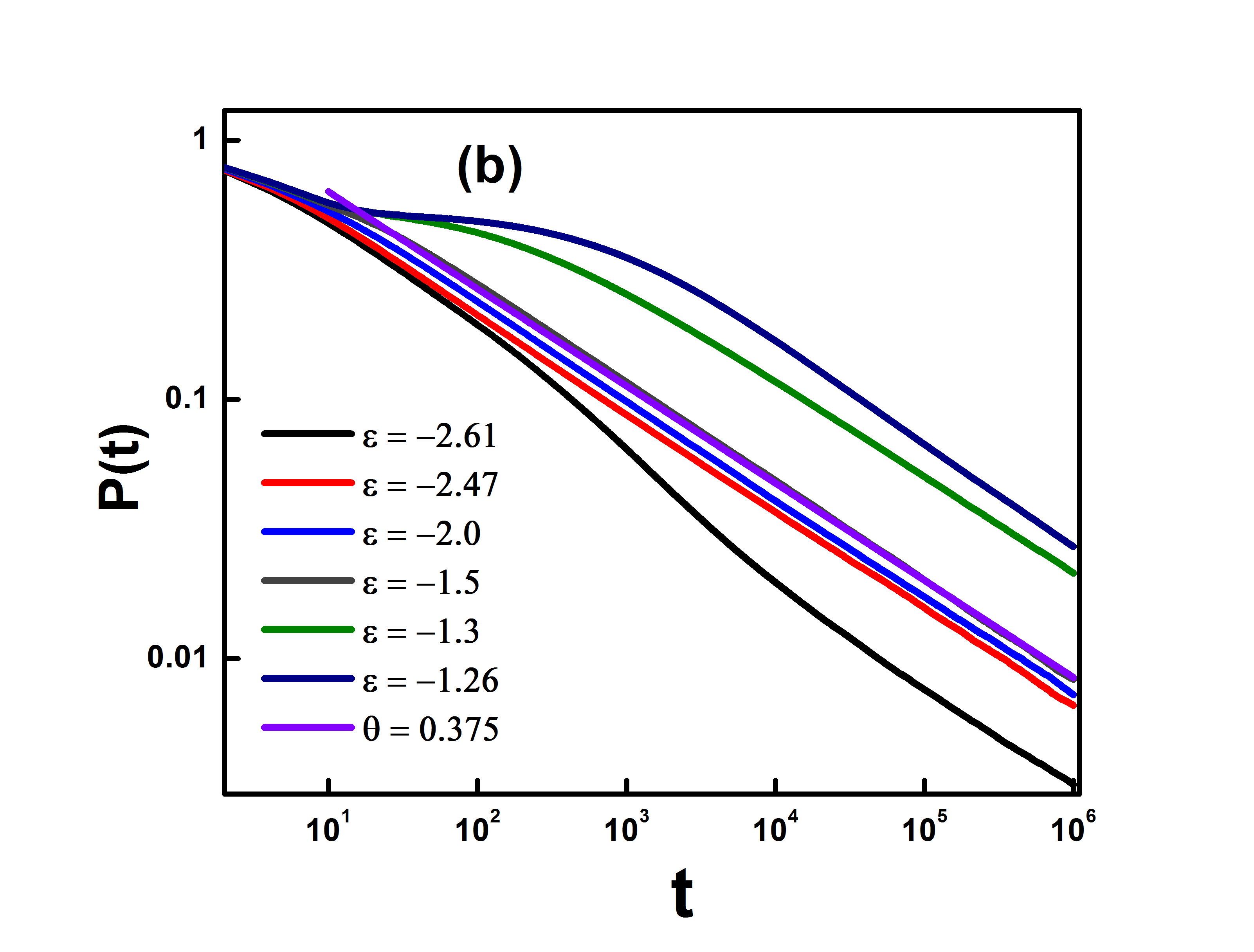}
       \caption{Coupled Gauss maps are simulated 
	for $N=3\times 10^5$ and averaged over a $20$ 
	configurations for values of coupling $\epsilon$
	such that $\epsilon_1<\epsilon<\epsilon_2$ where
	$\epsilon_1=-2.61$ and $\epsilon_2=-1.26$.
(a) We plot phase defect $D(t)$ as a function of time $t$ 
in the above $\epsilon$ range.  We observe, $D(t) \sim t^{-\delta}$ over this entire range and the decay exponent is $\delta=0.5$.
(b) We plot phase persistence $P(t)$ as a function of time $t$  
in this range. We note that 
$P(t) \sim t^{-\theta}$ with decay exponent of $\theta=0.375$.}
\label{fig3}
\end{figure*}

For Gauss map, we fix $\beta=0.8$ and $\nu=7.5$ in eq\ref{eq1} and 
compute both phase persistence $P(t)$ as well as phase defect 
$D(t)$ as a function of coupling $\epsilon$. 
We consider the 1-d lattice of size $N=3\times 10^5$. 
We simulate for $t=10^6$ time-steps and average over  $20$ configurations. 
We find that $D(t) \sim t^{-\delta}$ and $P(t) \sim t^{-\theta}$ over the range of 
critical parameter. This range is the same for both $P(t)$ and $D(t)$.
It ranges from $\epsilon=\epsilon_1=-2.61$ to
$\epsilon=\epsilon_2=-1.26$. We note that this range or spatial structure
cannot be inferred from the bifurcation diagram. In the entire range, 
the $D(t)$ has decay exponent is found to be $\delta=0.5$ while
$P(t)$ has decay exponent is $\theta=0.375$. The obtained values of
the persistent exponent are comparable with previously studied models 
such as the 1-d Ising model, coupled logistic maps and Sznajd model 
and the values of  $\delta=1/2$ are the same as that for the Ising class.
We have plotted  $D(t)$ in Fig.\ref{fig3}(a) and $P(t)$ in Fig.\ref{fig3}(b) 
for various values of $\epsilon$ such that $\epsilon_1<\epsilon<\epsilon_2$. 

\begin{figure*}[h]
       \includegraphics[scale=0.3]{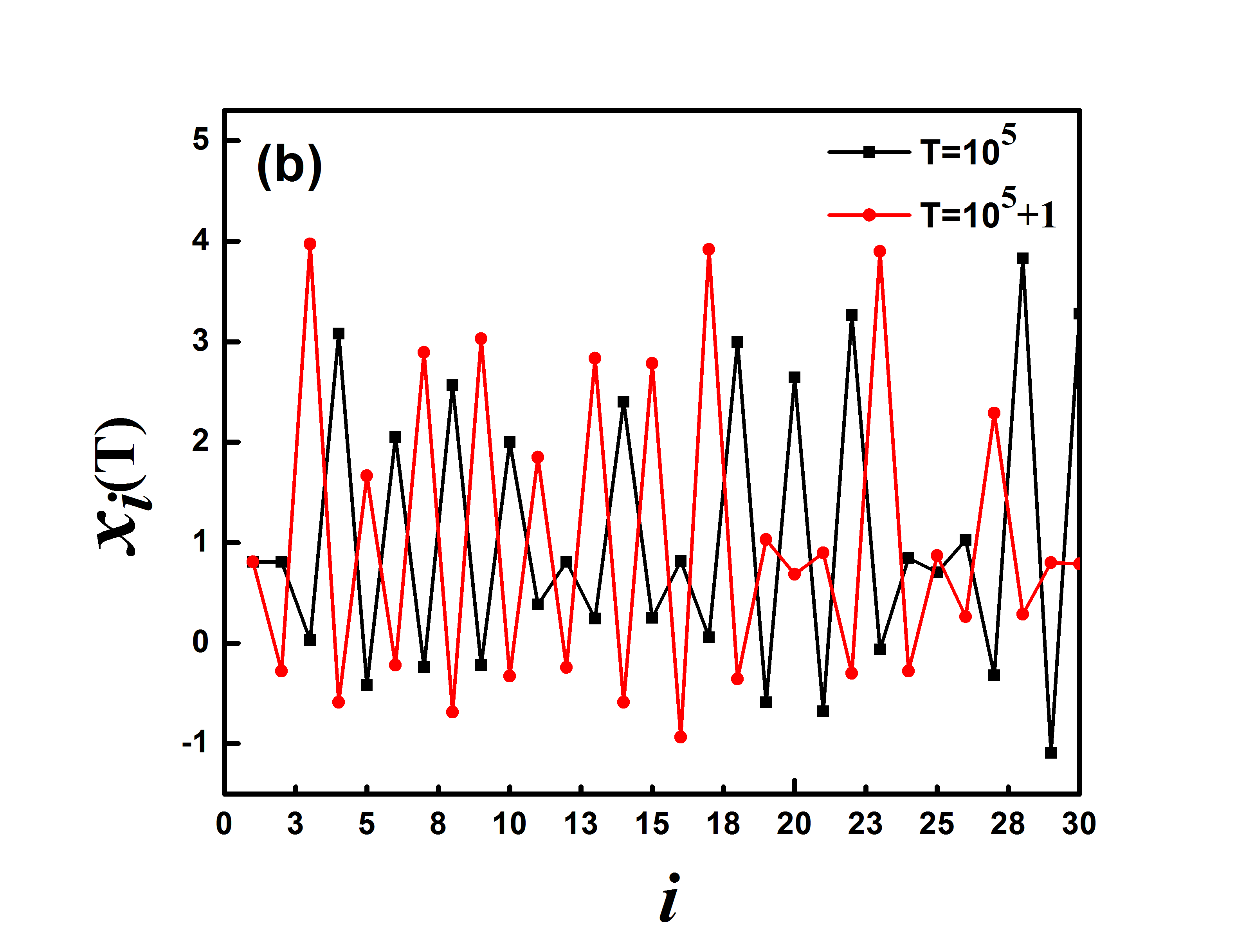}
       \includegraphics[scale=0.3]{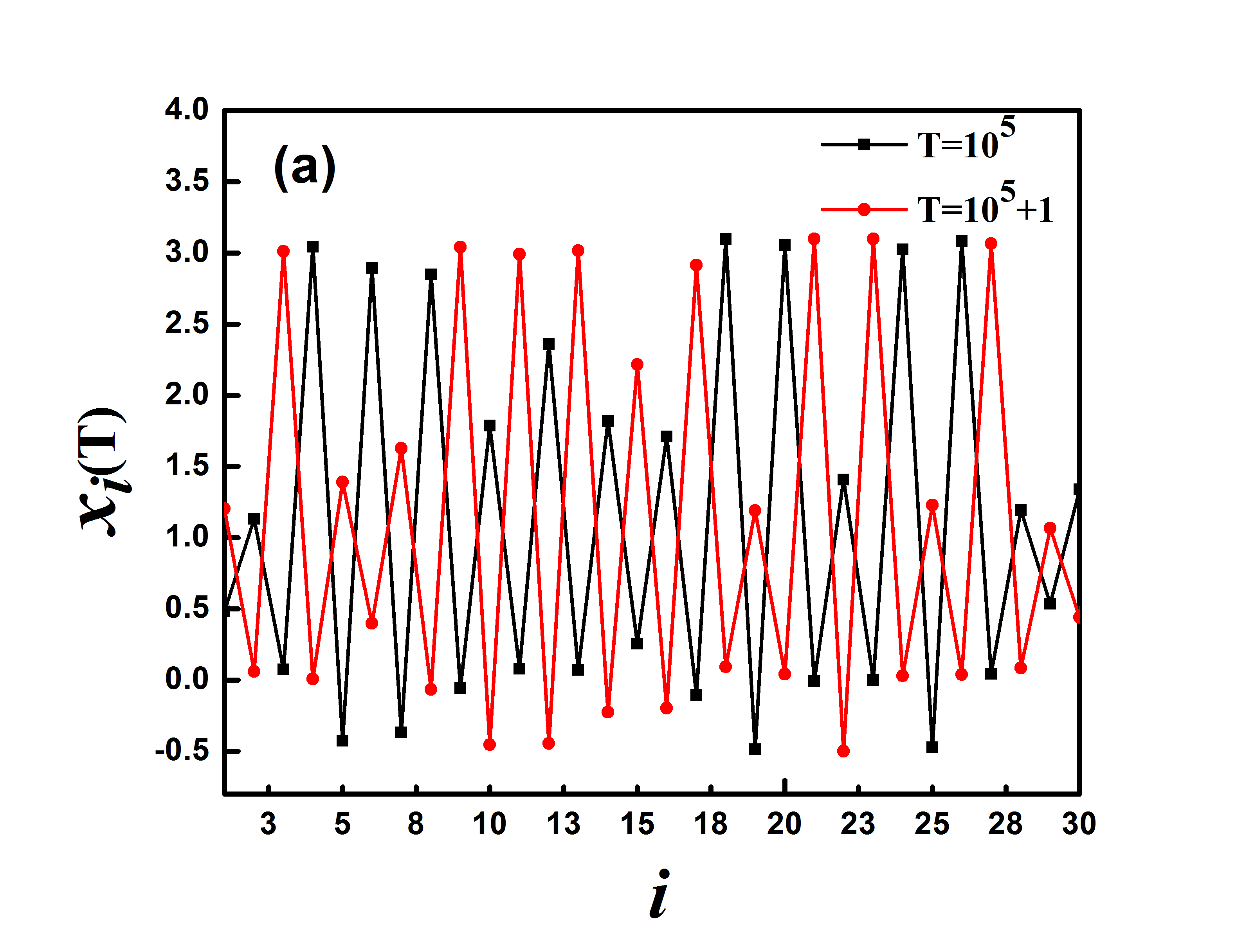}
       \caption{ We plot spatial profile $x_{i}(T)$ versus $i$
	fort $T=10^5$, 
	and $T=10^5+1$ for 
	a) $\epsilon=-2.2$ b) $\epsilon=-1.3$.
	The lattice size is $N=100$. However,
	only 30 sites are plotted for clarity.}
       \label{fig4}
\end{figure*}

We plot the spatial profile $x_{i}(T)$  as a function of
site index $i$(See Fig. \ref{fig4}). As expected,
the zigzag pattern is reached within the critical
parameter region. We simulate $N=100$ sites and show
its state at $T=10^5$ and $T=10^5+1$
for $\epsilon=-2.2$ and $\epsilon=-1.3$. 
An up-down pattern at even time steps becomes down-up
pattern at odd time steps. This does not affect
phase defects. However, since the spin values are
regained after even time steps, we compute persistence
only for even time steps.

\begin{figure*}[h]
       \includegraphics[scale=0.3]{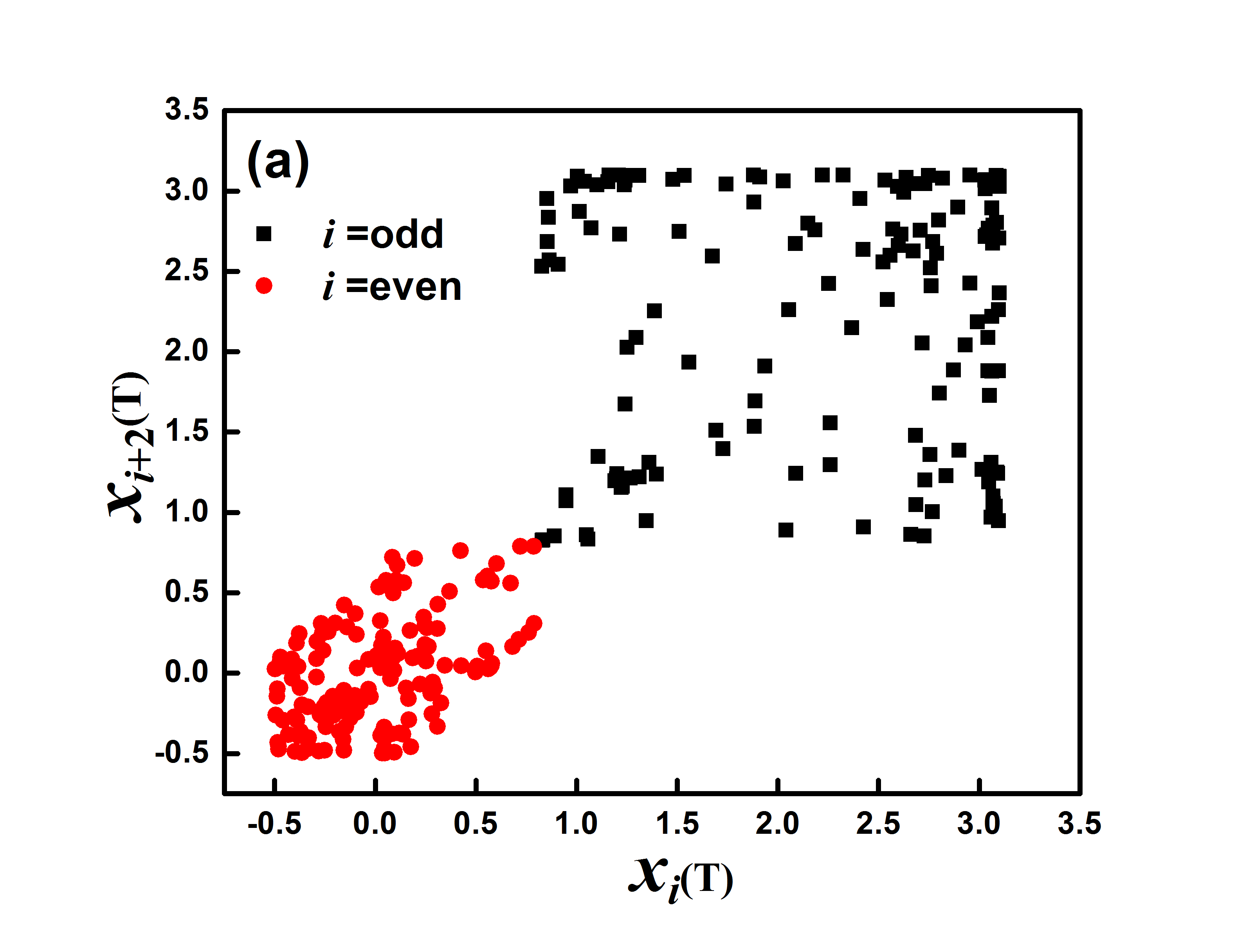}
       \includegraphics[scale=0.3]{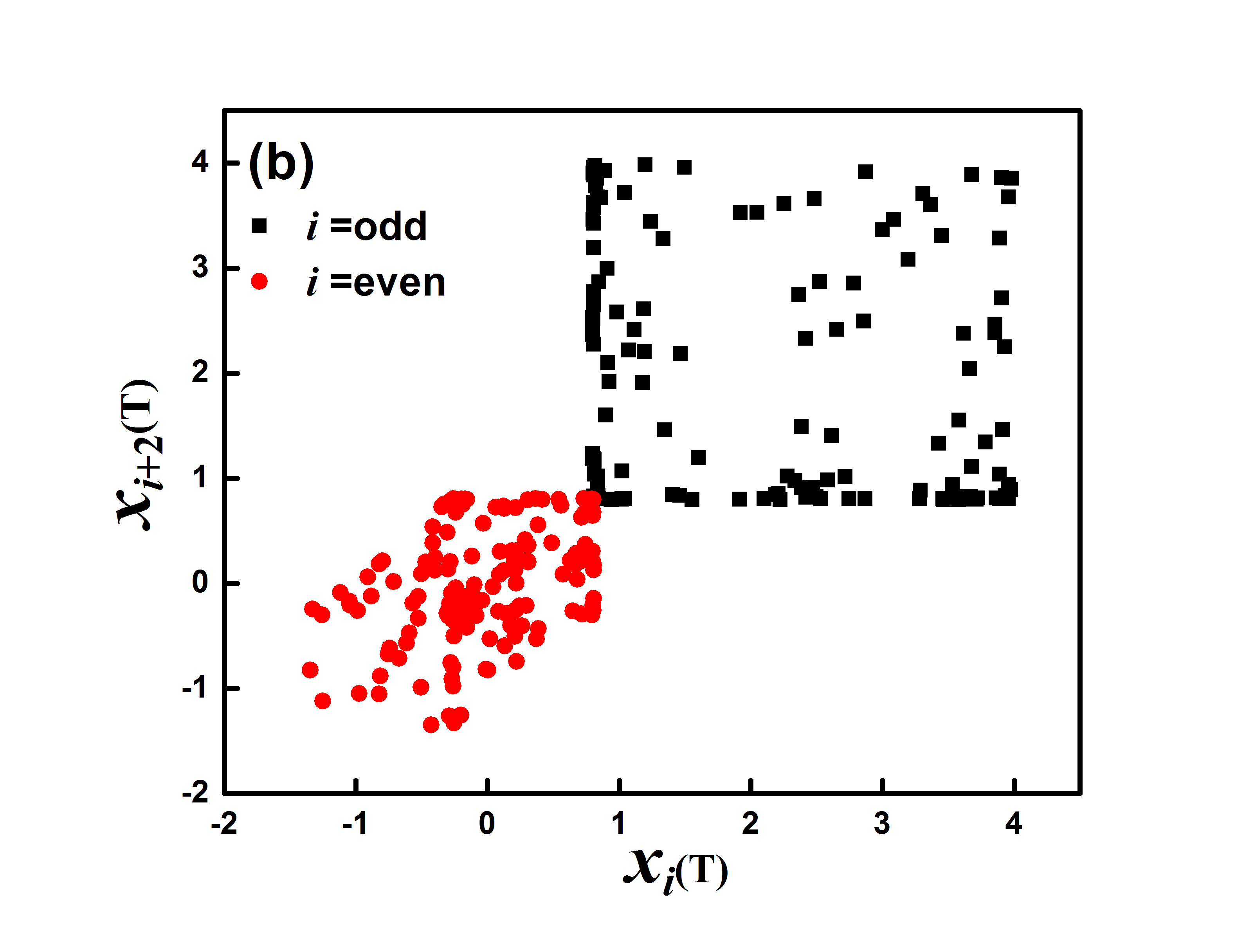}
	\caption{We plot spatial return map $x_{i+2}(T)$ versus $x_{i}(T)$
        for both even as well odd sites. We consider $N=300$ 
	and $T=2\times10^5$
        and simulate for a) $\epsilon=-1.3$, b) $\epsilon=-2.2$. }
       \label{fig5}
\end{figure*}

We have plotted spatial return map $x_{i+2}(T)$ versus $x_{i}(T)$
for both even as well odd sites 
for the same values of $\epsilon$, {\it{i.e}} 
$\epsilon=-1.3$ and $\epsilon=-2.2$ 
(See Fig. \ref{fig5})
for which we have given spatial profiles.
Spatial return map plotted for $N=300$ after time $T=2 \times 10^5$.
There is no synchronous chaos. The values are in two different bands.
There may or may not be any separation between bands. Thus the pattern is more complex than synchronous chaos. Since there may not be
any separation between bands, the bifurcation diagram 
does not help to identify this phase.

\begin{figure*}[h]
       \includegraphics[scale=0.3]{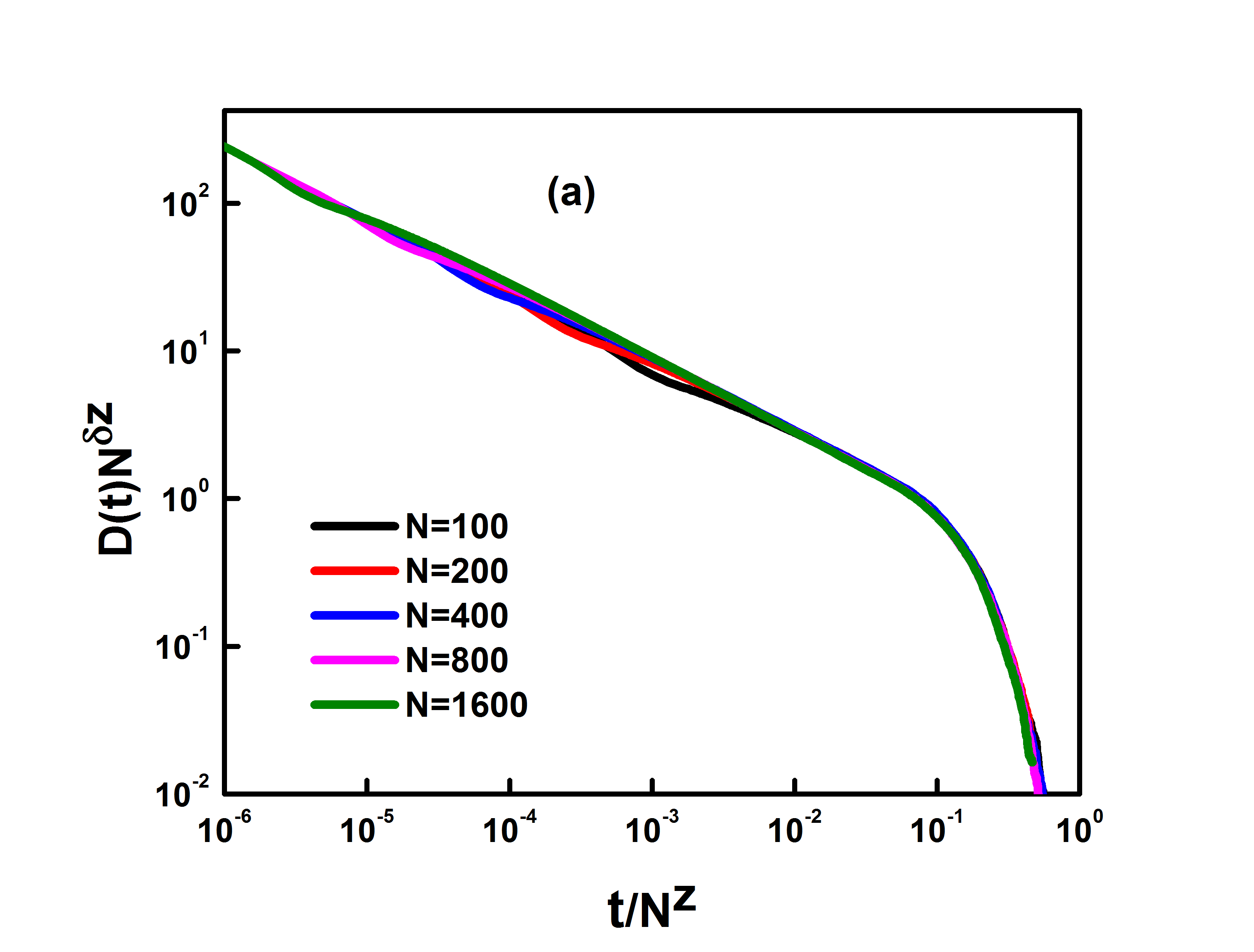}
       \includegraphics[scale=0.3]{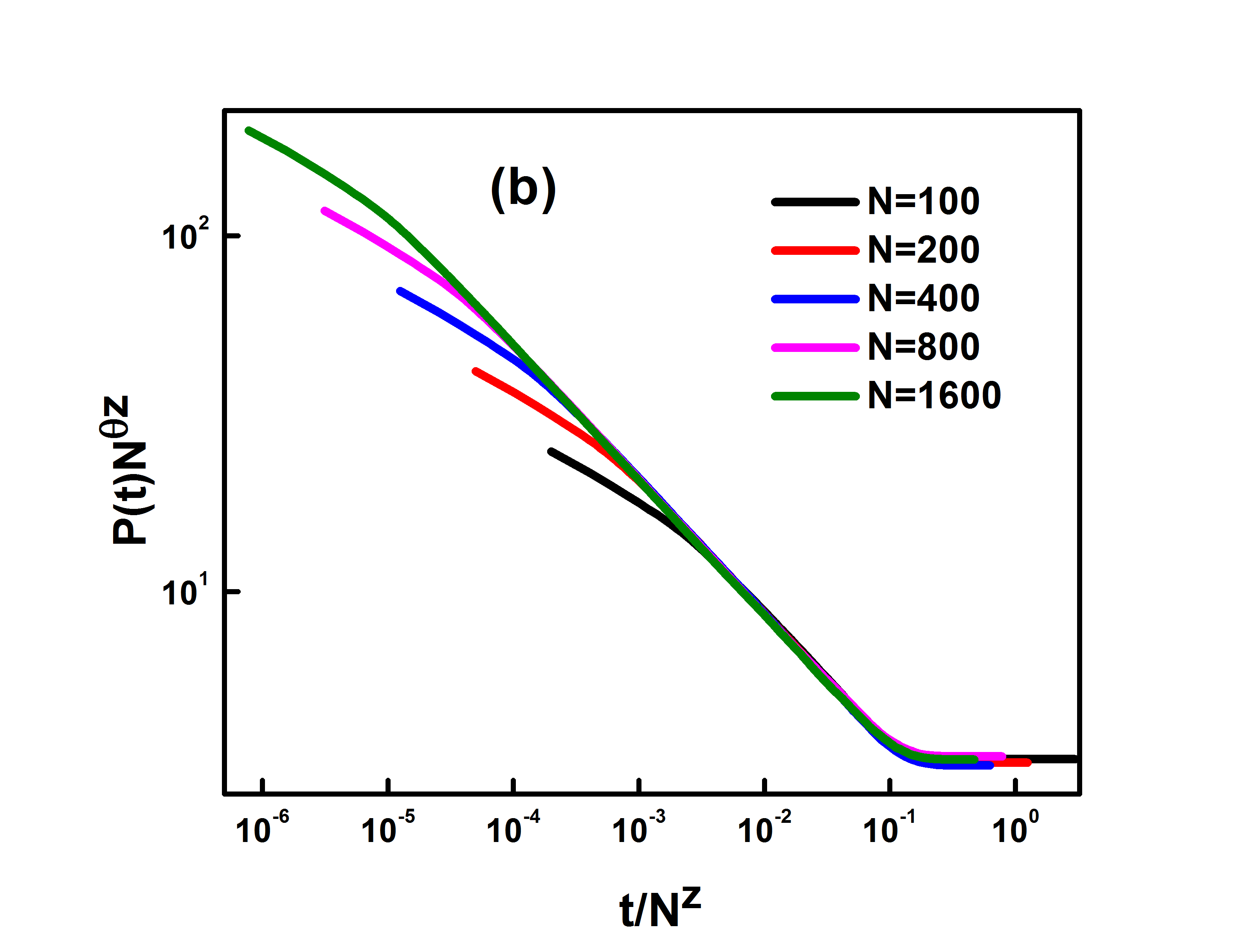}
       \caption{a) $D(t)N^{\delta z}$ is plotted as a
function of $t/N^z$ for $N=100,200,400,800,1600$ at $\epsilon=-1.5$.
A good scaling collapse is observed at $z=2$ 
b)  $P(t)N^{\theta z}$ is plotted as a function of $t/N^z$ for $N=100,200,400,800,1600$ at $\epsilon=-1.5$.
A good scaling collapse is observed at $z=2$.}
\label{fig6}
\end{figure*}

We carry out finite-size scaling to compute the dynamic exponent $z$.
We consider $N=100,200,400,800,1600$  for  
a value of $\epsilon$ in range $\epsilon_1<\epsilon<\epsilon_2$.
We consider $\epsilon=-1.5$ and each averaged over a $5000$ 
configurations. We have plotted $D(t)N^{\delta z}$ as a function of
$t/N^z$ (see Fig.\ref{fig6}(a)). A fine collapse is obtained at $z=2$. 
In Fig.\ref{fig6}(b), we plot $P(t)N^{\theta z}$ as a function of $t/N^z$.
The excellent scaling collapse is obtained at $z=2$.

\begin{figure*}[h]
       \centering
       \includegraphics[scale=0.35]{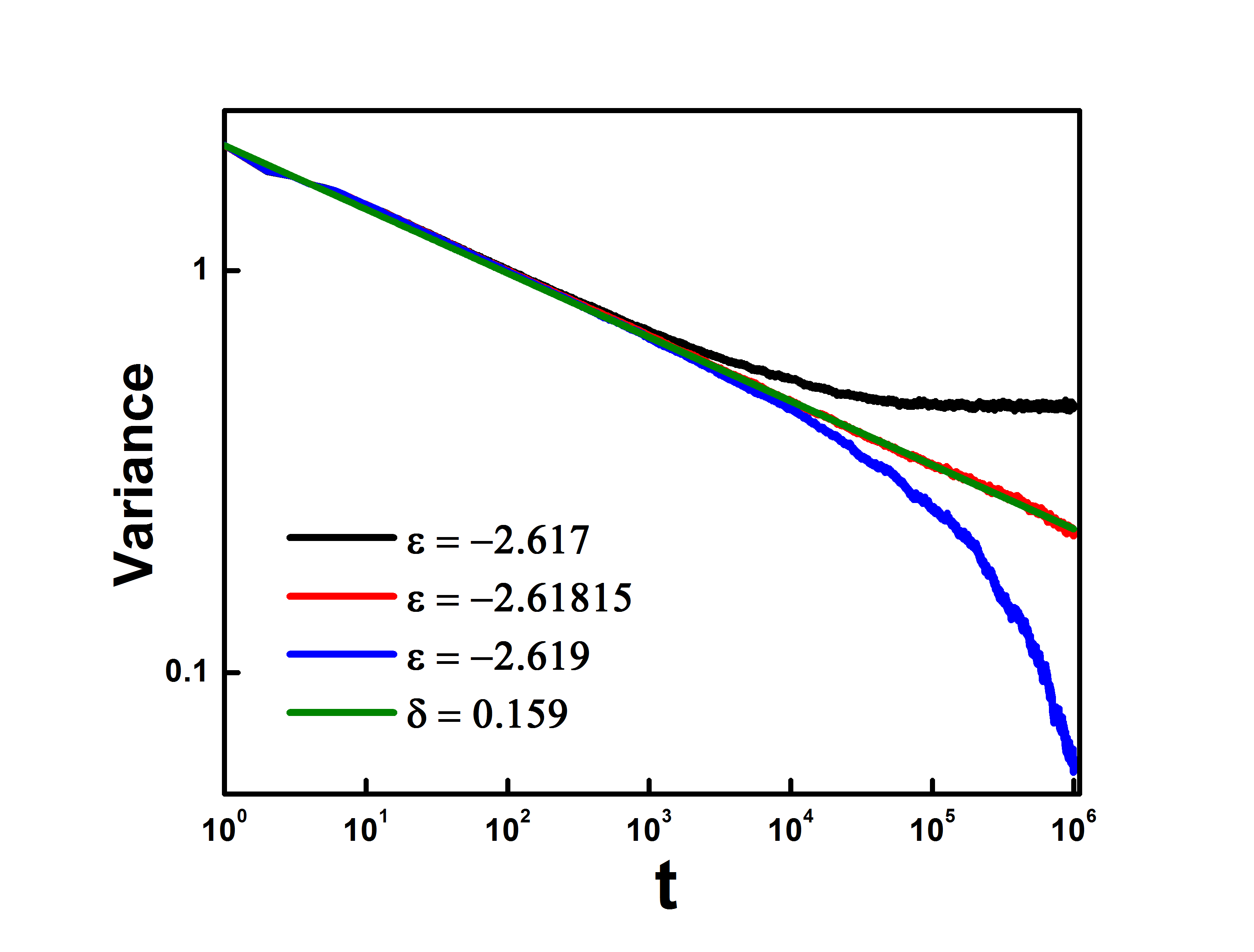}
       \caption{Variance is plotted as a function of time for synchronized 
        values of $\epsilon$. 
	Power-law decay is observed at $\epsilon=-2.61815$ with $\delta=0.159$. The lattice size is $N=10^5$.
	We average over a $50$ configuration.}
       \label{fig7}
\end{figure*}

We observe synchronization at certain critical coupling $\epsilon_c$. 
For completeness,
we study the transition to synchronization using variance as 
an order parameter. Here the transition is in directed percolation (DP) class.
We consider $N=10^5$ and average over a $50$ configuration. We plot variance as a function of time for the transition to
a synchronized state at $\epsilon_c$. It decays as a power-law in time 
at $\epsilon_c=-2.61815$. 
The power-law exponent is found to be $\delta=0.159$ (See Fig. \ref{fig7}) which matches with directed percolation.
This transition is on the expected lines.

\begin{figure*}[h]
       \includegraphics[scale=0.3]{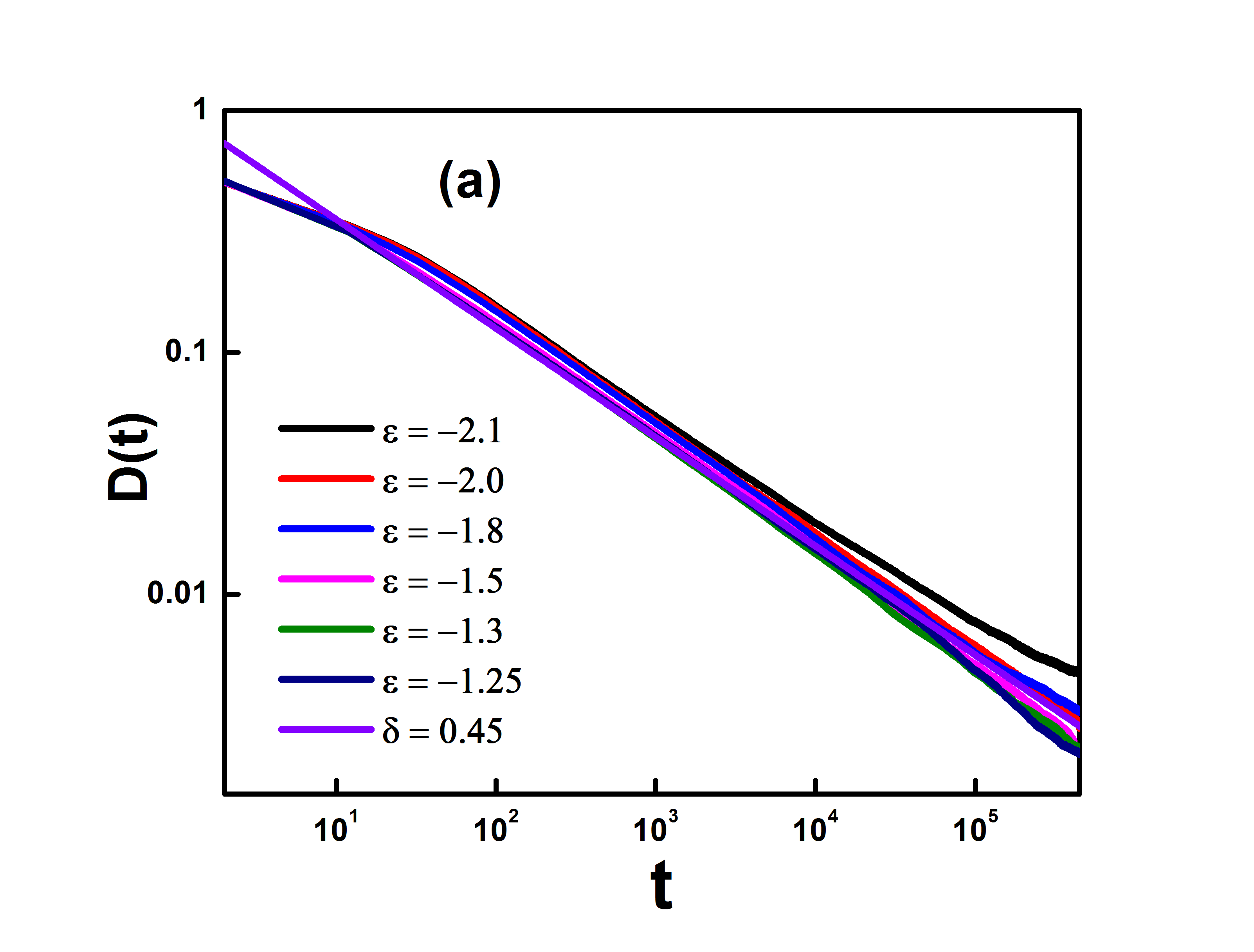}
       \includegraphics[scale=0.3]{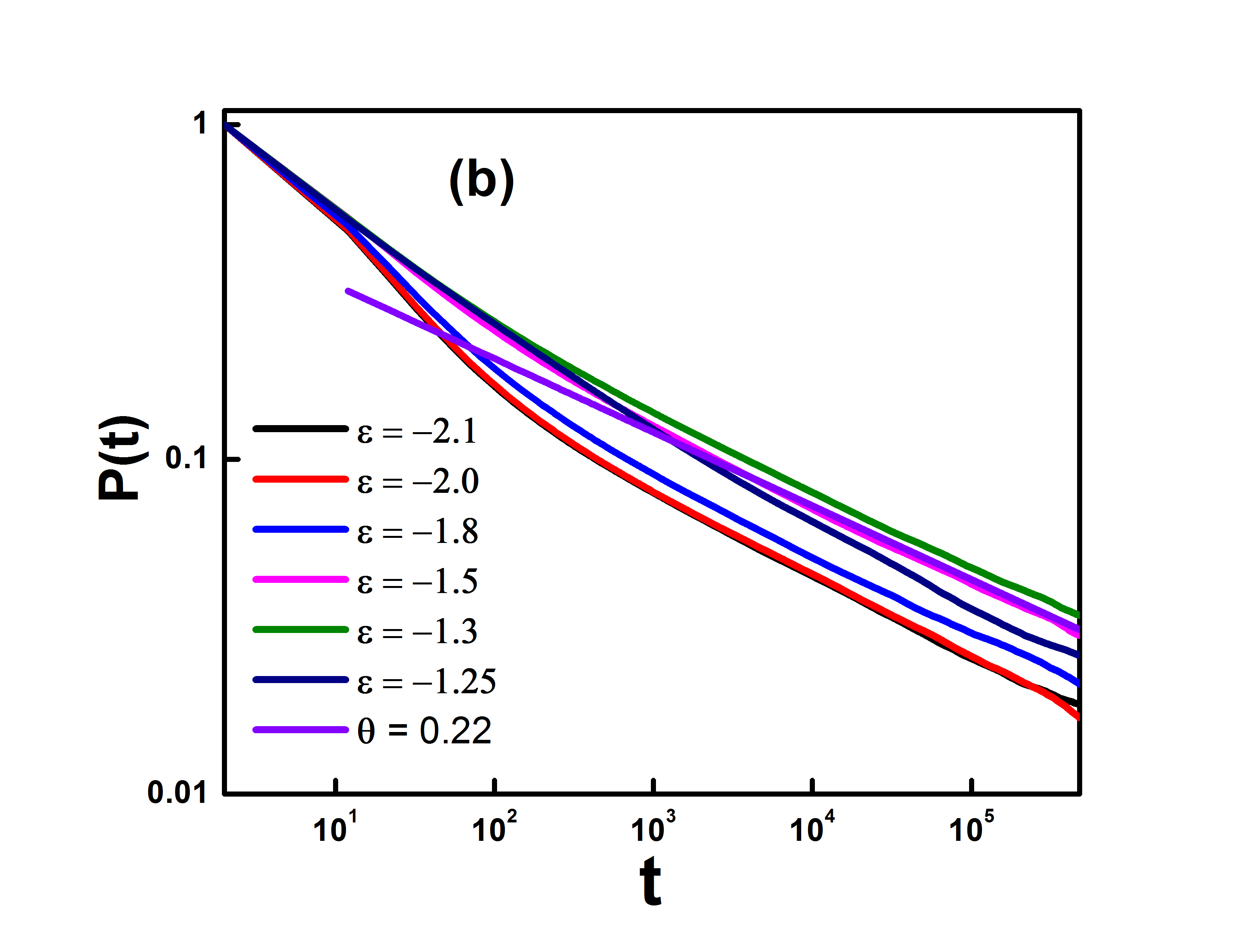}
       \caption{For coupled Gauss maps in two dimensions,
        we plot phase defect $D(t)$ and phase persistence $P(t)$ as
        a function of time $t$. We consider $10^3\times 10^3$ lattice
        and average over $10$  configurations in the range
	$\epsilon_1<\epsilon<\epsilon_2$.  ($\epsilon_1=-2.1$ and
	$\epsilon_2=-1.25$).
	a) $D(t)$ shows power-law
        decay with exponent $\delta=0.45$.
        b) $P(t)$ shows power-law  decay with exponent $\theta=0.22$.}
       \label{fig8}
\end{figure*}

We also study the Gauss map on two-dimensional lattice for
$\nu=9$ and $\beta=-0.5$ in eq.\ref{eq1}. 
We simulate a 2-d lattice for $N=10^3$ and average
over $10$ configurations. Similar to 1-d lattice, 
we find that $D(t) \sim t^{-\delta}$ and $P(t) \sim t^{-\theta}$ over a large range of parameters. 
This range is the same for both $D(t)$ and $P(t)$. It ranges from $\epsilon=\epsilon_1=-2.1$
to $\epsilon=\epsilon_2=-1.25$. In this range, 
$D(t)$ decays with the exponent of $\delta=0.45$ and $P(t)$ decays with an exponent of $\theta=0.22$. 
We plot $D(t)$ in Fig.\ref{fig8}(a) and  $P(t)$ in Fig.\ref{fig8}(b) for various values
of $\epsilon$ such that $\epsilon_1 < \epsilon < \epsilon_2$.

\begin{figure*}[h]
       \includegraphics[scale=0.3]{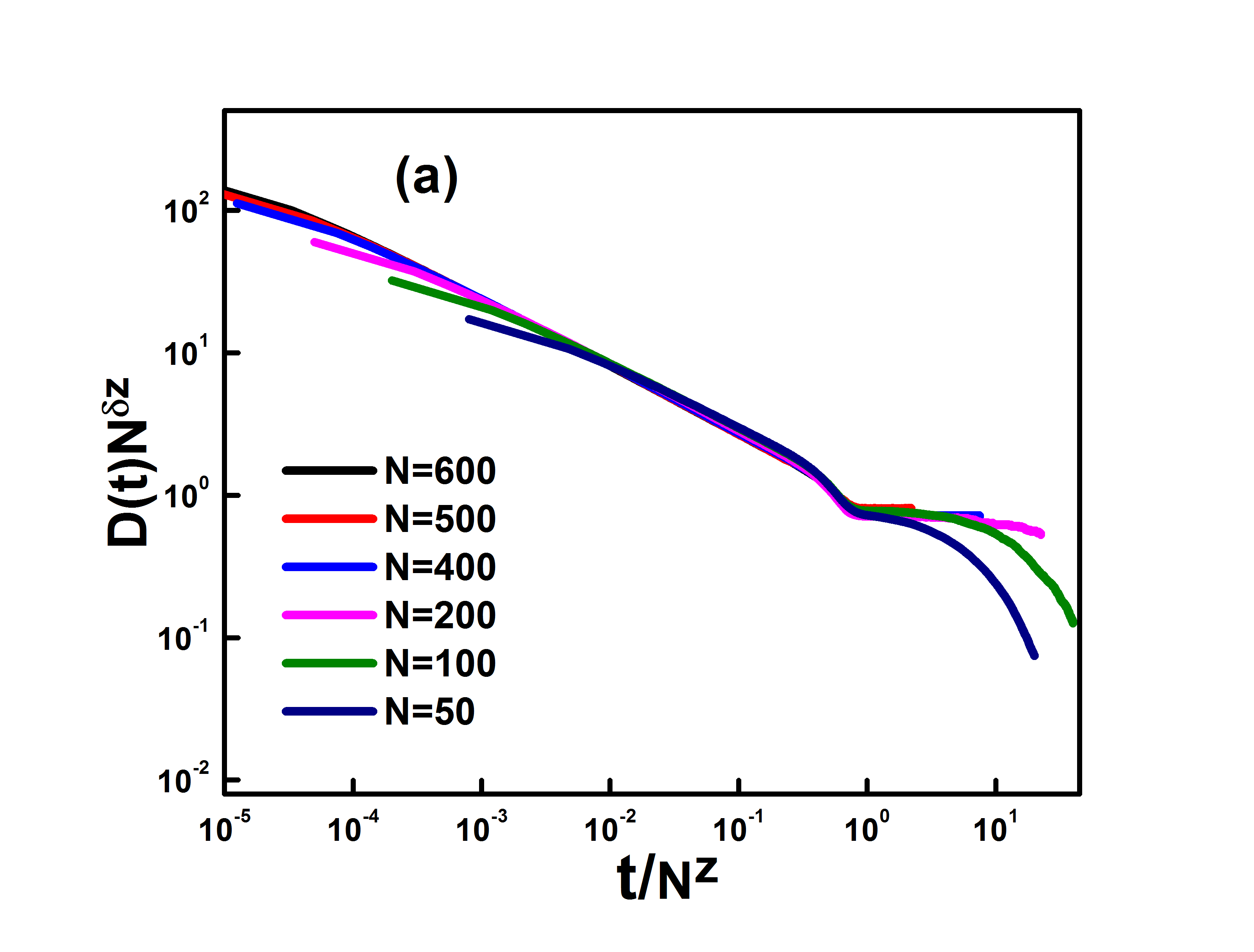}
       \includegraphics[scale=0.3]{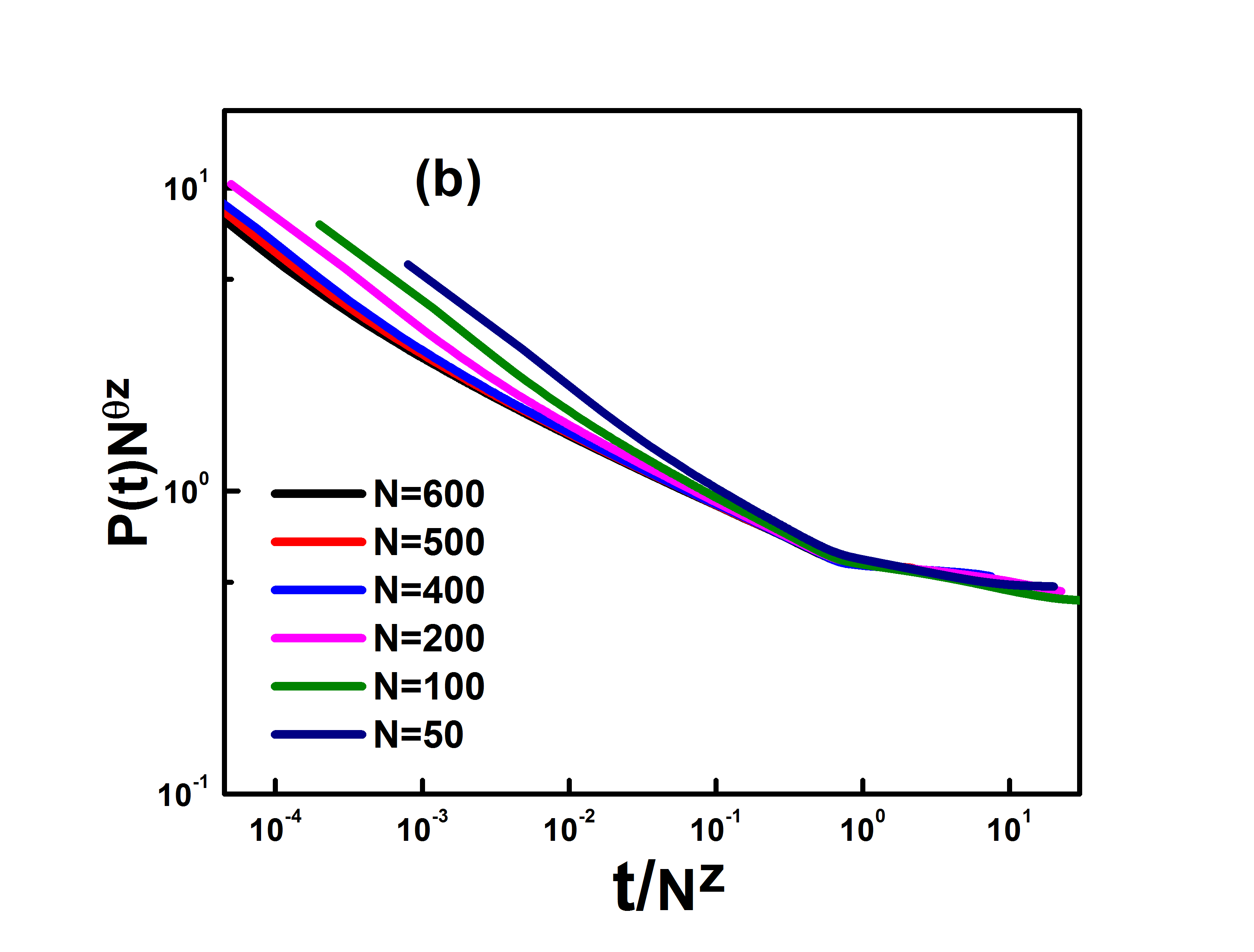}
       \caption{a) $D(t)N^{\delta z}$ is plotted as a function
of $t/N^z$ for $N=$ 600, 500, 400, 200, 100, 50 at $\epsilon=-1.5$.
	A good scaling collapse is observed at $z=2$. 
	b) $P(t)N^{\theta z}$  is plotted as a function of $t/N^z$ for 
$N=$ 600, 500, 400, 200, 100, 50 at $\epsilon=-1.5$. We ignore long-lived metastable states after the first kink. A good scaling collapse is observed at $z=2$.}
       \label{fig9}
\end{figure*}

We carry out the finite-size scaling for the 2-d lattice as well.
We simulate the $N\times N$ lattice for $N=$ 600, 500, 400, 200, 100, 50
and average more than $4 \times 10^3$ configurations for $N \le 200$ and 500 configurations for $N > 200 $. 
We observe that the power law is followed by a metastable state which decays to zero.
Thus there are two kinks.
We  plot $D(t)N^{\delta z}$ as a function of $t/N^z$ at $\epsilon_c=-1.5$.
The fine collapse is obtained at $z=2$  for the first kink(See Fig.\ref{fig9}(a)). Similarly, we also plot $P(t)N^{\theta z}$ as a function
of $t/N^z$. The fine collapse is obtained at $z=2$ for the first 
kink(See Fig.\ref{fig9}(b)). It is different
dynamics after the first kink. Such two-stage coarsening dynamics is also obtained previously in models of opinion dynamics \cite{PhysRevE.102.012316}. This departure of $z$ from the expected value of $2.16$ for the Ising model may be due to a long-lived metastable state at long times.

We also look at the time evolution of the states. 
The time evolution of spins is similar to  non-conserved
dynamics of Ising model known as model A\cite{hohenberg1977theory}. 
The coarsening is similar to what is observed for quenching at zero
temperature. Finally, a single giant phase of a zigzag or checkerboard pattern is obtained.

\begin{figure*}[h]
       \centering
       \includegraphics[scale=0.3]{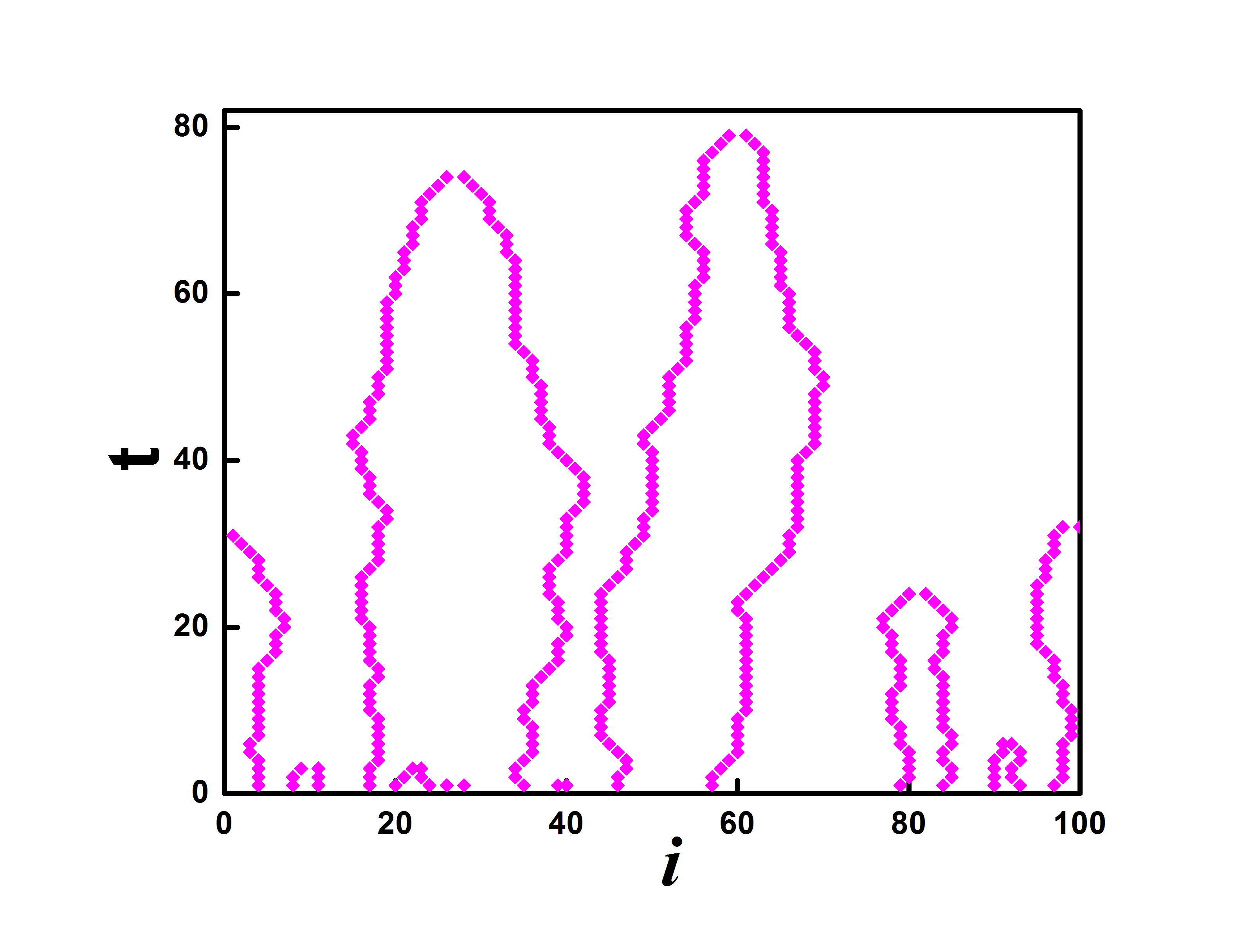}
       \caption{We present spatiotemporal dynamics of defects at
	$\epsilon$ = -1.5 for coupled Gauss maps in one dimension.}
       \label{fig10}
\end{figure*}

In one dimension, we visualize dynamics of defects {\it{i.e.}} consecutive
sites with the same spin.
In Fig.\ref{fig10} we plot the position of defects in time. We 
observe that they carry out random walks, annihilate each other and
we obtain lattice with effective antiferromagnetic ordering.  

\begin{figure*}[h]
       \includegraphics[scale=0.195]{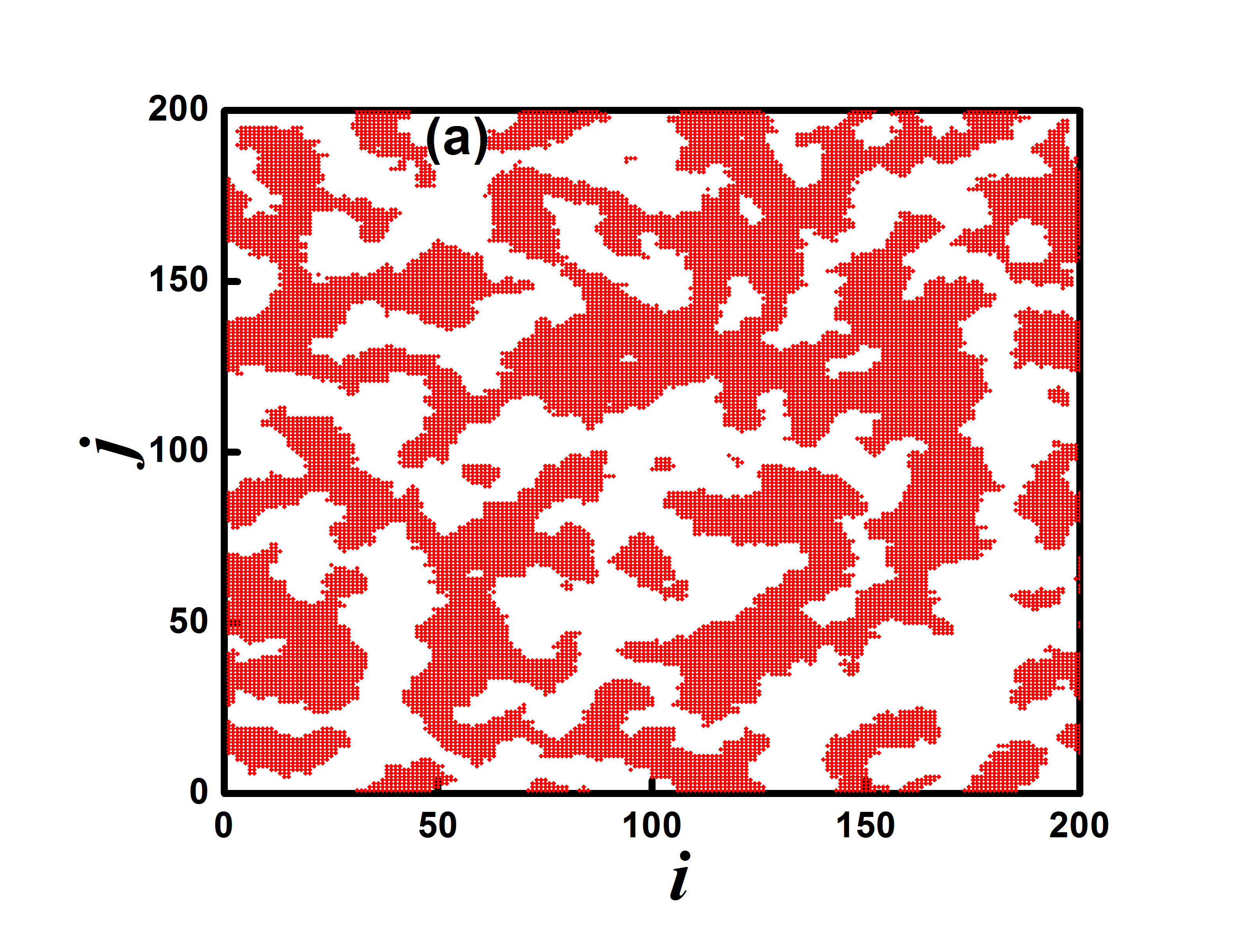}
       \includegraphics[scale=0.195]{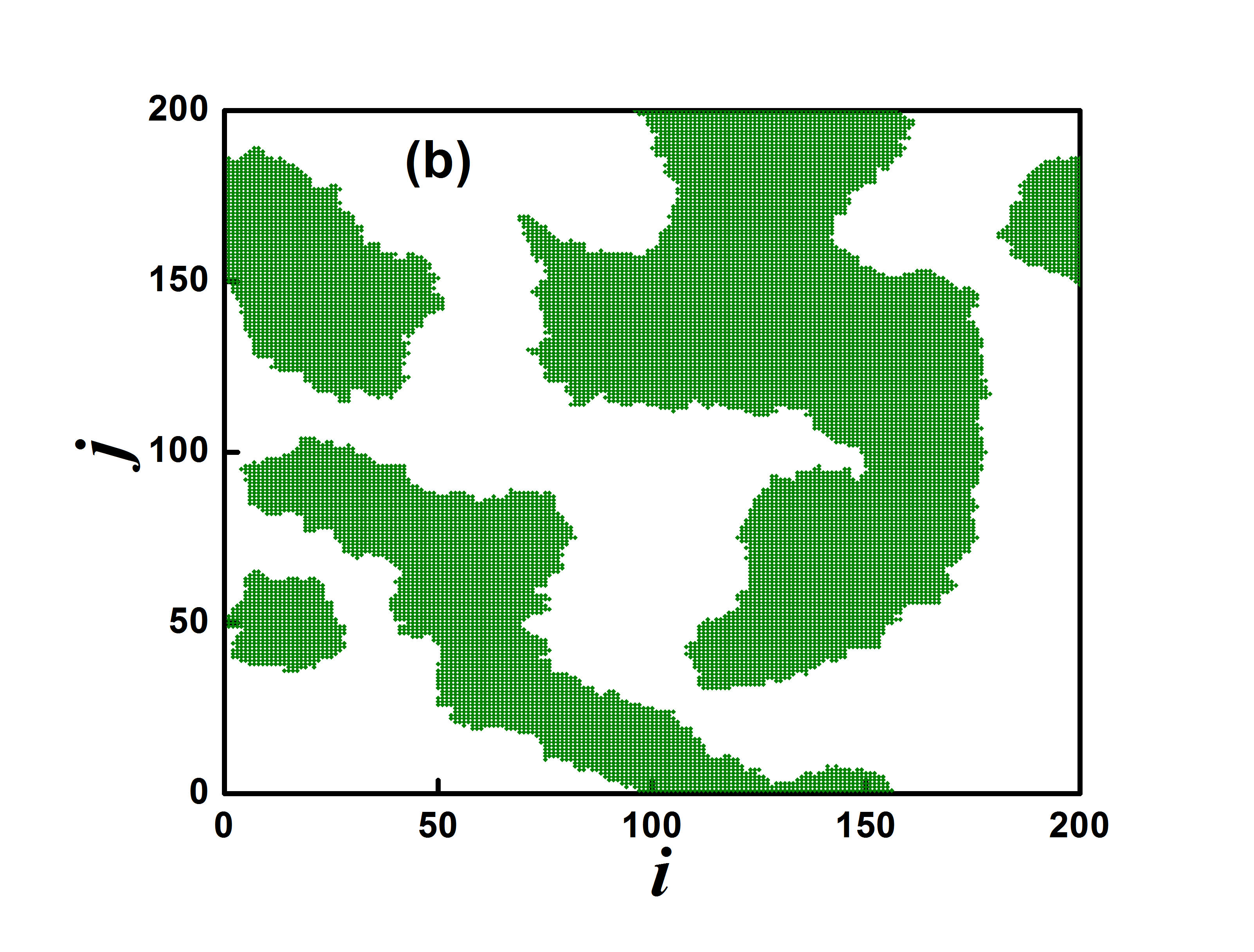}
       \includegraphics[scale=0.195]{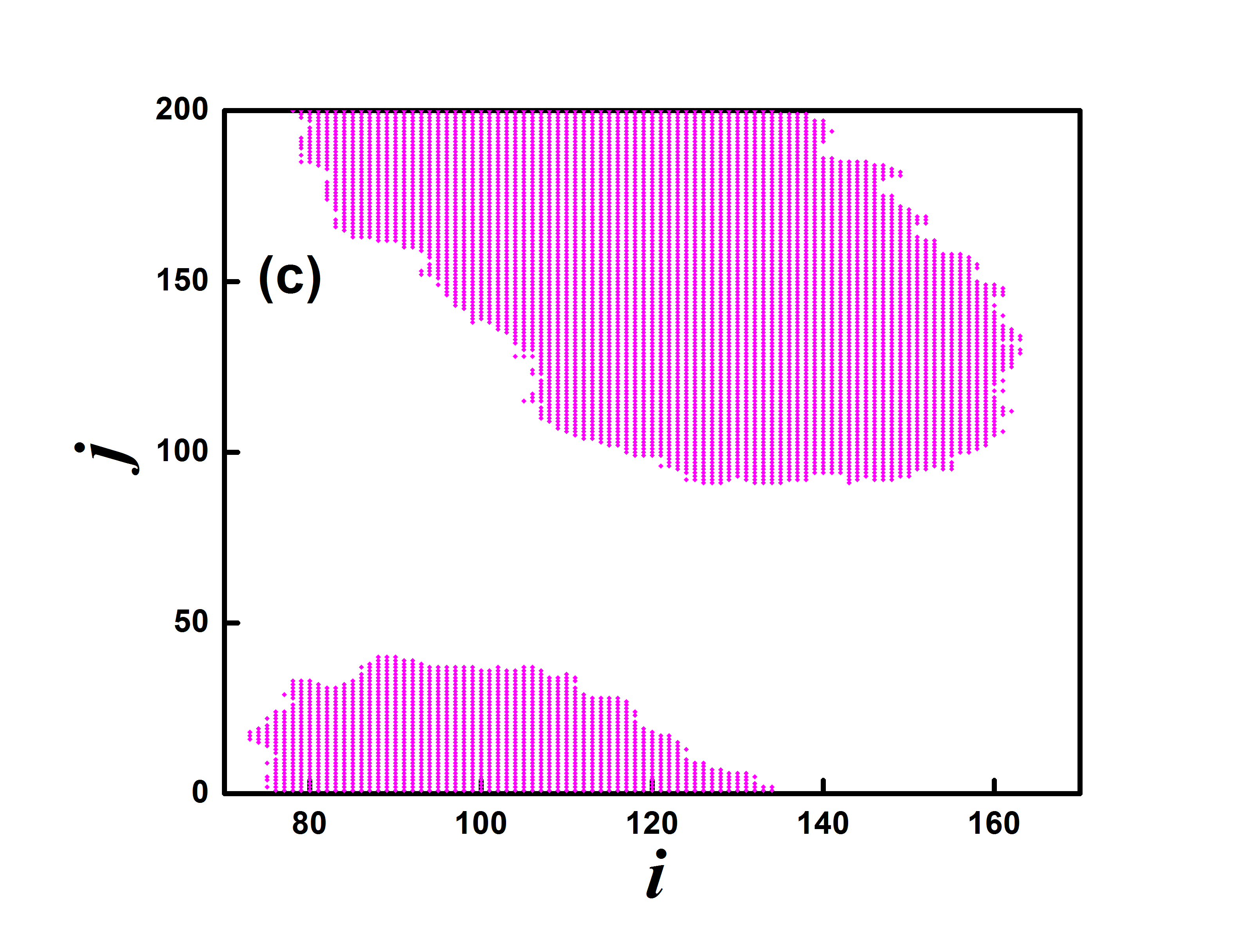}
       \caption{For Gauss map, we plot sites $(i,j)$ such that
	$(-1)^{i+j} s(i,j) = 1$ for 
	a)  $t$ = $10^2$ , b)  $t$ = $10^3$, and 
	c)  $t$ = $10^4$ }
       \label{fig11}
\end{figure*}

To visualize the phase in two dimensions, we change the
sign of the odd sublattice
of the square lattice. The checkerboard pattern becomes
a ferromagnetic pattern in this representation. We consider $N=200$, $\epsilon=-1.5$ and plot sites $(i,j)$ such that $(-1)^{i+j} 
s_{i,j}=1$. The extent of the checkerboard pattern (ferromagnetic pattern 
in this representation) grows in time till it spans the entire
lattice. Three such representations 
in increasing time are plotted in Fig.\ref{fig11}.

\subsection{Logistic map}

\begin{figure*}[h]
       \includegraphics[scale=0.3]{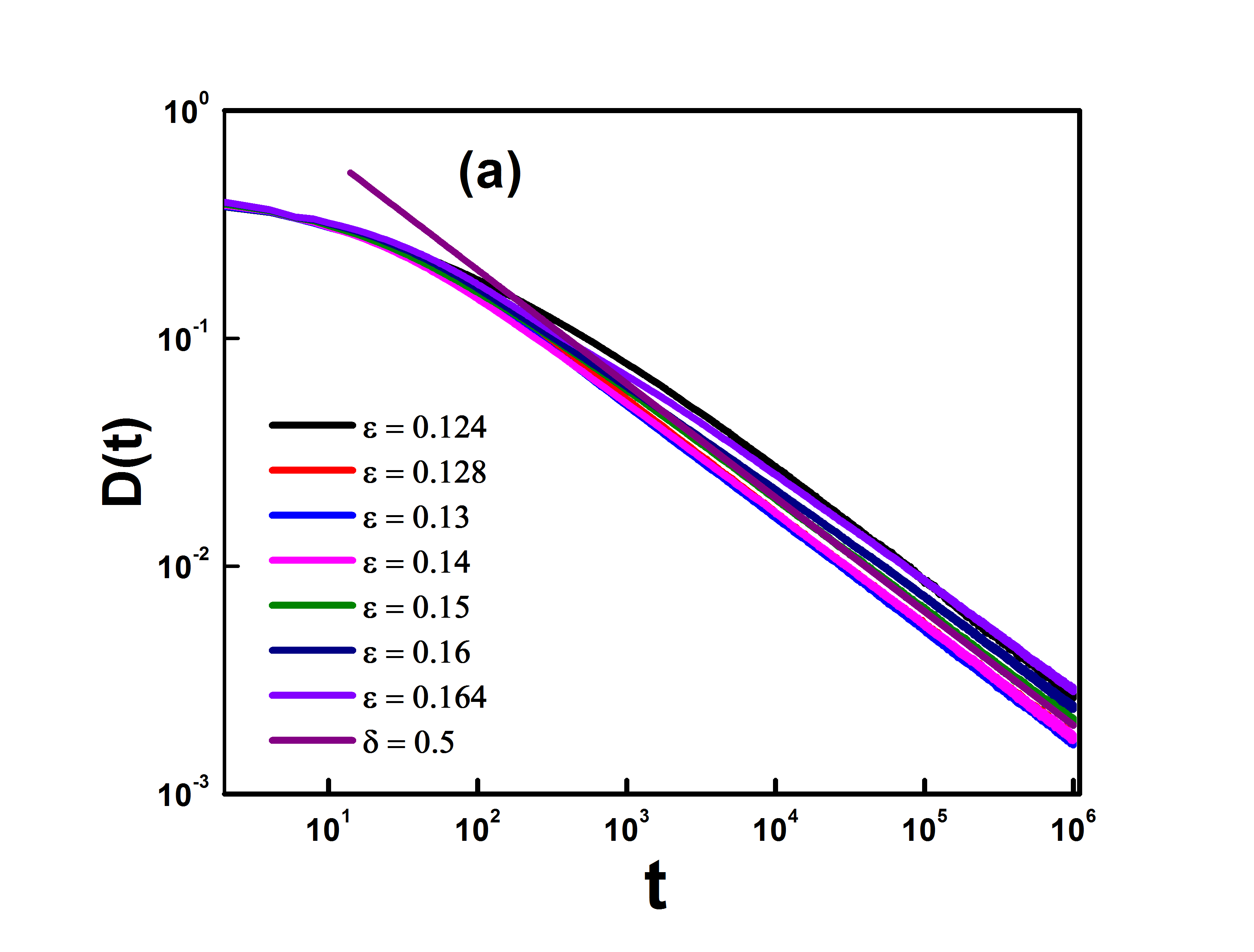}
       \includegraphics[scale=0.3]{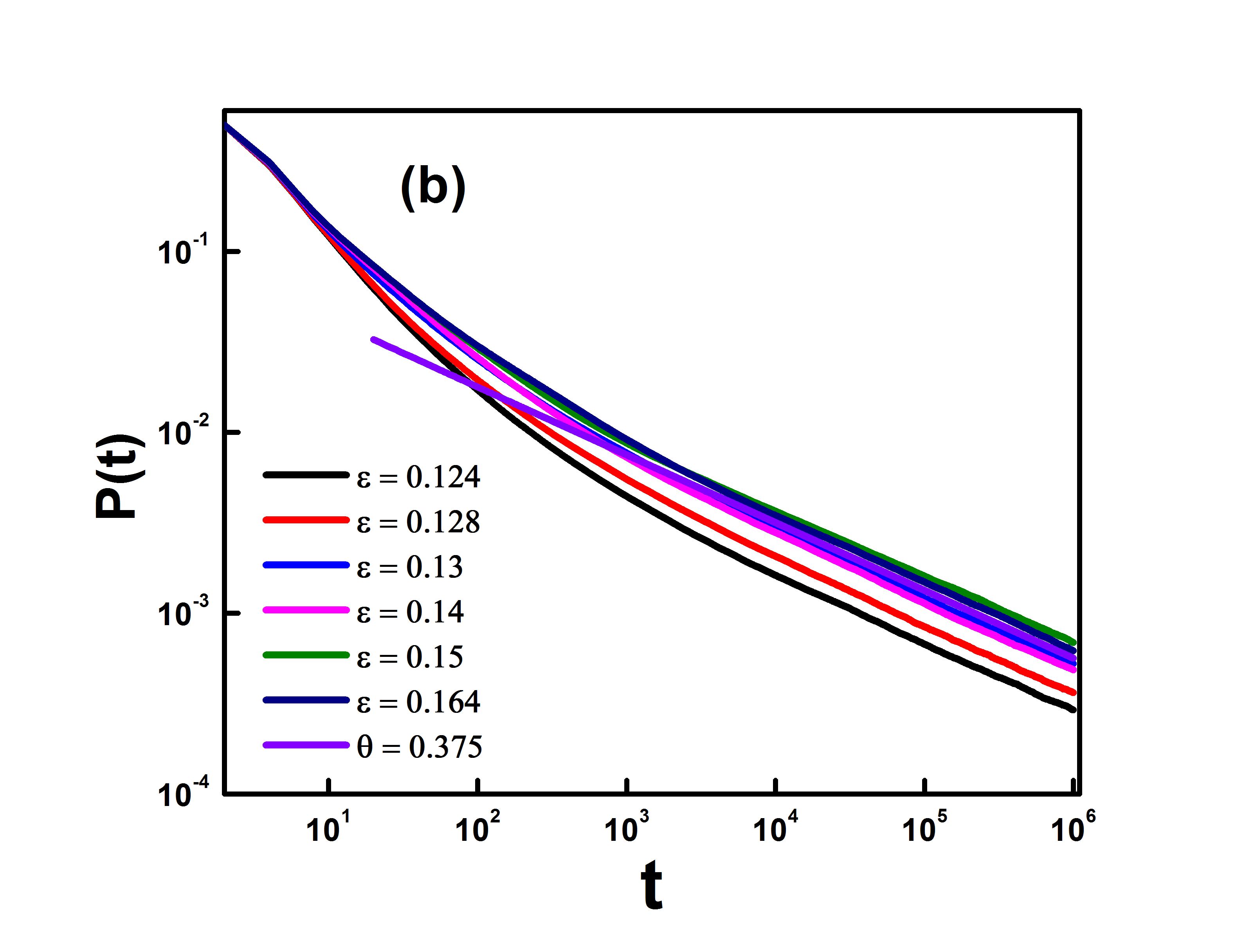}
       \caption{We show the evolution of phase defect
	$D(t)$ and phase persistence $P(t)$
	as a function of time $t$ for coupled logistic map
	for $\epsilon_1 < \epsilon < \epsilon_2$ where $\epsilon_1=0.124$ and 
	$\epsilon_2=0.164$. The lattice size is   $N=3\times 10^5$ and we average over a $20$ configuration.
	a) $D(t)$ as function of time $t$. 
	b) $P(t)$ as a function of time $t$.
	It is clear that both $D(t)$ and $P(t)$ decay as power-law with 
	exponent $0.5$ and $0.375$ respectively throughout above range. }
       \label{fig12}
\end{figure*}

To investigate how generic these observations are, we study the
system of the coupled logistic map. We consider the system
described by eq. \ref{eq3} in one dimension and eq. \ref{eq4} in
two dimensions.
We fix $\mu=4$ in eq.\ref{eq2} and vary the coupling parameter $\epsilon$.  In one dimension, we consider a lattice of size $ N=3\times 10^5$,  simulate for $t= 10^6$ timesteps and average over a $20$ configuration. 
Again, we observe power-law decay of phase defect
$D(t)$  and phase persistence  $P(t)$ over a range of coupling
$\epsilon_1=0.124< \epsilon<\epsilon_2=0.164$.
In this entire range $D(t)\sim t^{-\delta}$ and
$P(t)\sim t^{-\theta}$ with $\delta=0.5$ and $\theta=0.375$. 
These specific values of exponents
are the same as those observed for the Gauss map. 
We have demonstrated this power-law behaviour by plotting $D(t)$
in Fig.\ref{fig12}(a) and $P(t)$ in Fig.\ref{fig12}(b) 
for various values of $\epsilon$ such that $\epsilon_1 < \epsilon < \epsilon_2$

\begin{figure*}[h]
       \centering
       \includegraphics[scale=0.35]{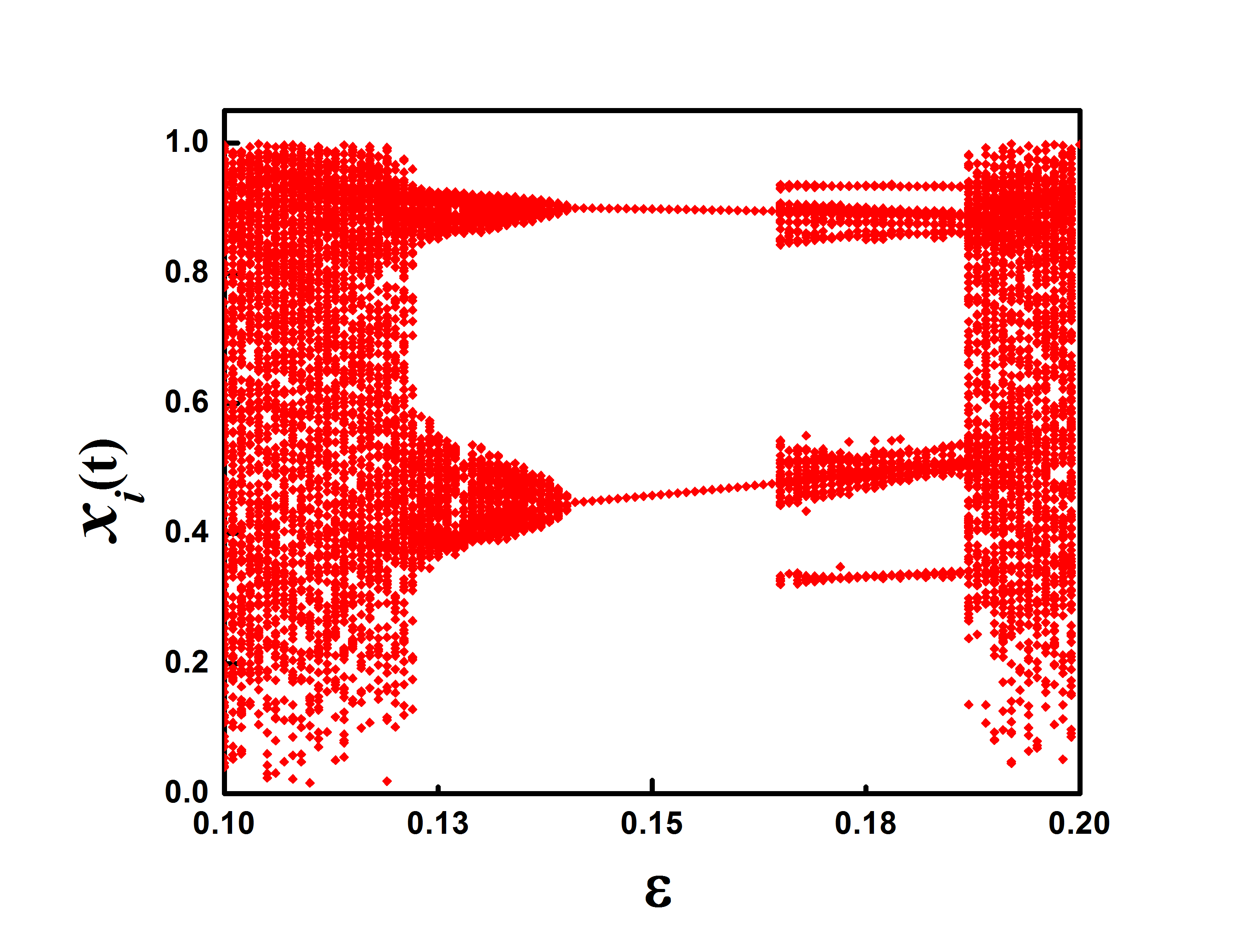}
       \caption{Bifurcation diagram of coupled logistic map for $\mu=4$.
	We plot all sites $x_{i}(t)$ as a function
        coupling $\epsilon$ at $t=5.8 \times 10^5$ and $N=200$.}
	\label{fig13}
\end{figure*}

We also show a bifurcation diagram for 
coupled logistic maps.
We consider lattice size $N=200$ and simulate 
for $t=5.8 \times 10^5$ time steps. 
We find that two bands are observed over $0.124<\epsilon<0.186$. 
However, a zigzag pattern is observed only over
the range $0.124<\epsilon<0.164$ (See Fig.\ref{fig13})
Thus the presence of two bands in the bifurcation diagram is neither necessary nor sufficient for the observation of zigzag patterns.
It is also clear that this state cannot be inferred from bifurcation
diagram.

\begin{figure*}[h]
       \includegraphics[scale=0.195]{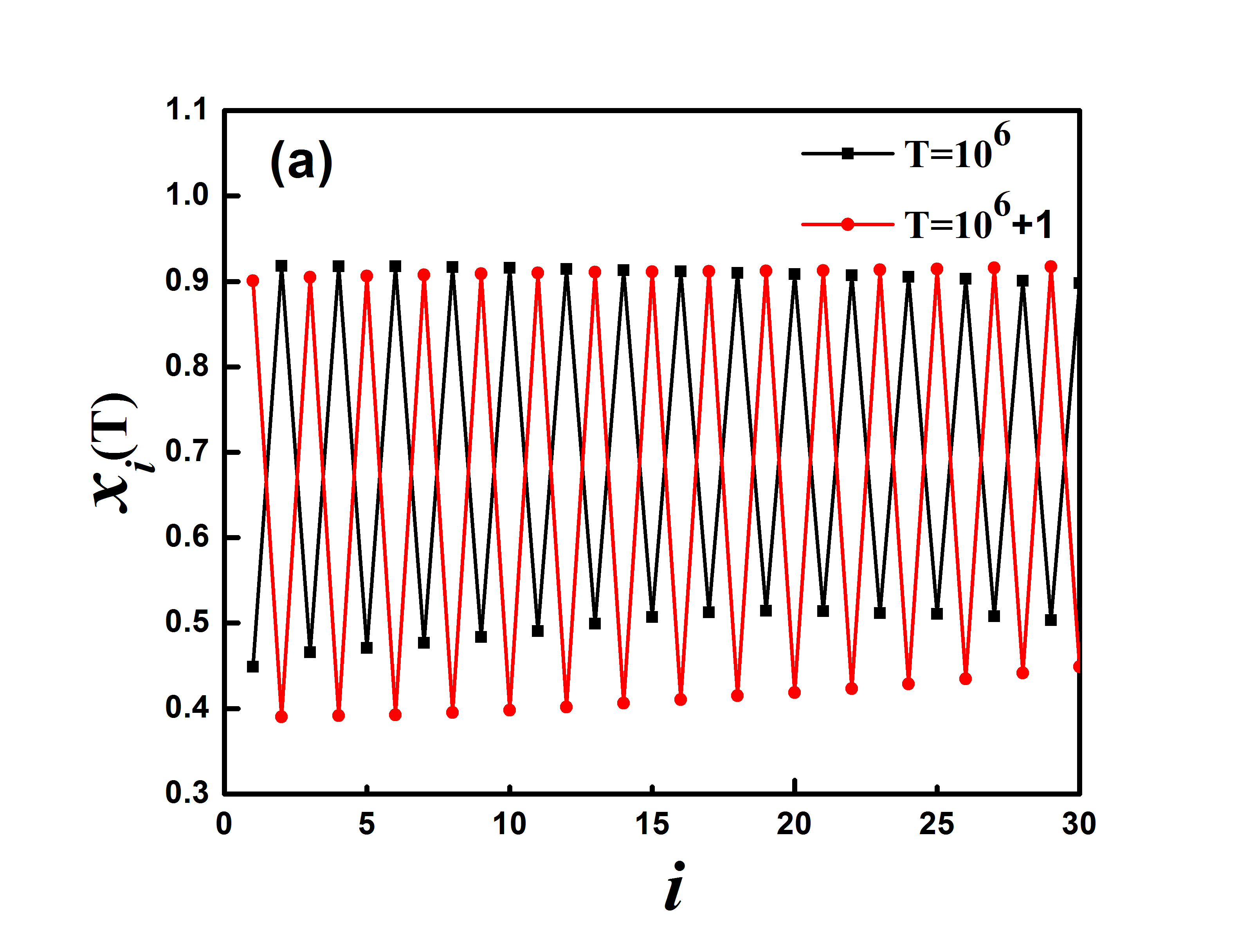}
       \includegraphics[scale=0.195]{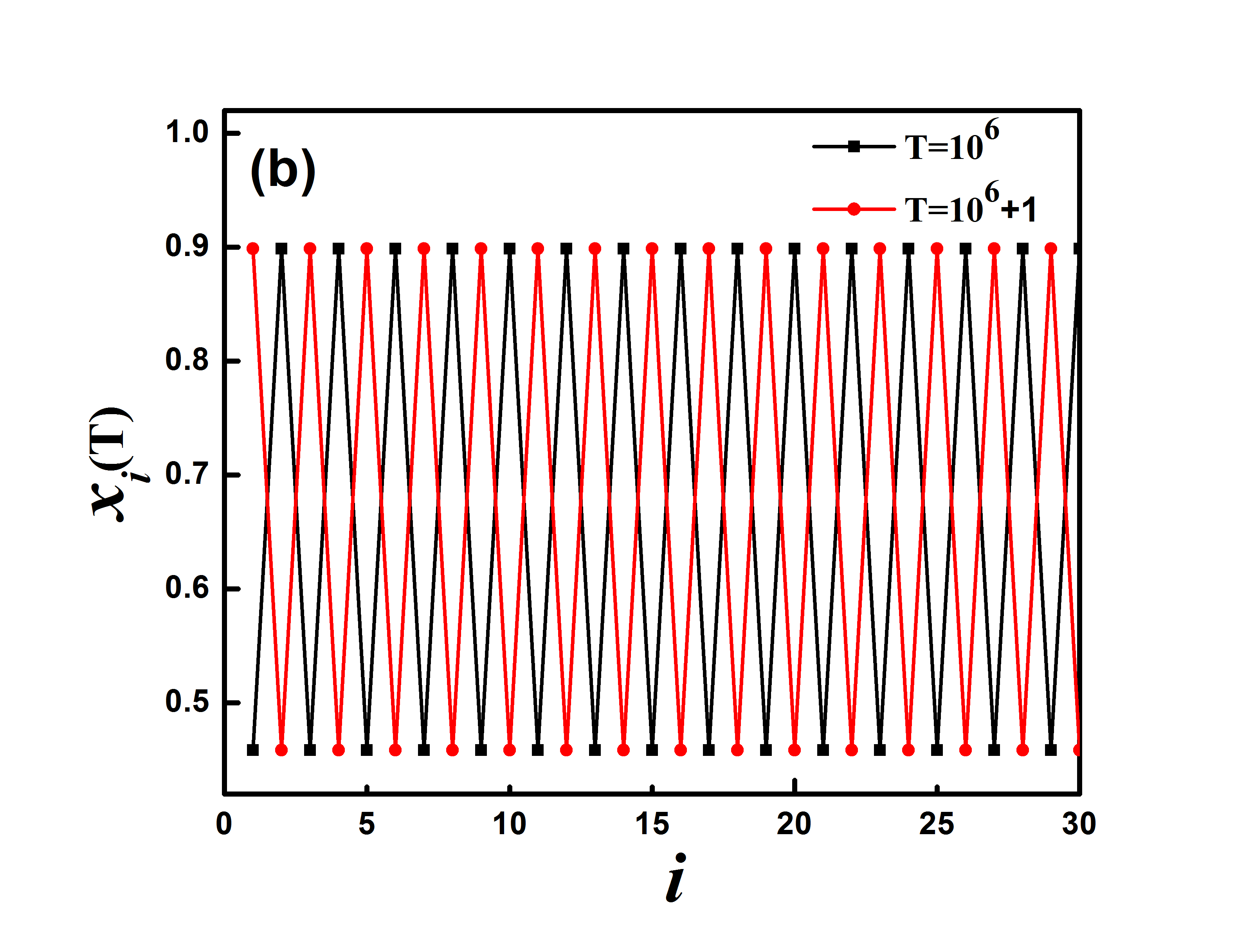}
       \includegraphics[scale=0.195]{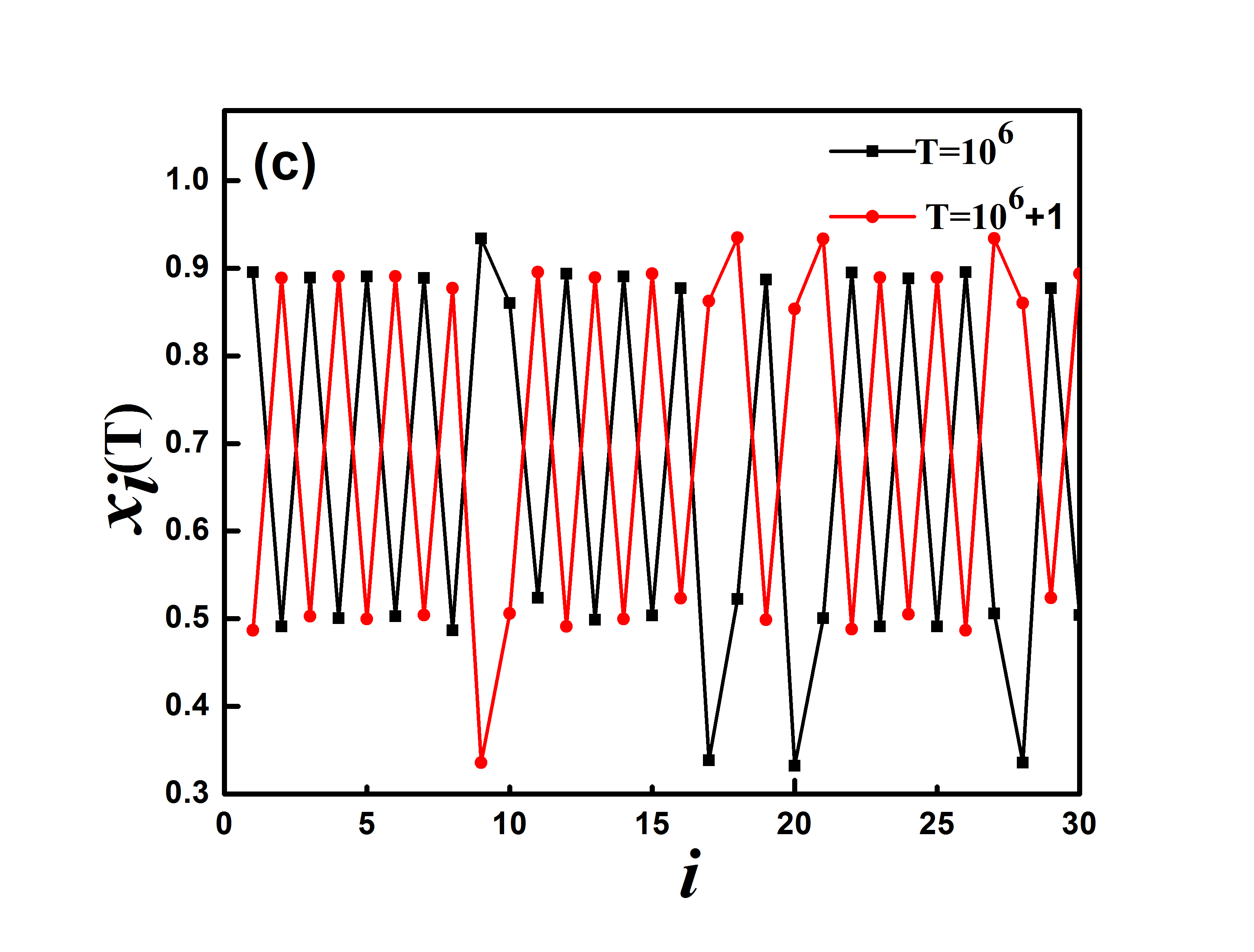}
       \caption{ We plot spatial profile $x_{i}(T)$ versus $i$
	for time steps $T=10^6$ 
	and $T=10^6+1$ and size $N=100$ for
	a) $\epsilon=0.13$, b) $\epsilon=0.15$ c) $\epsilon=0.18$ }
       \label{fig14}
\end{figure*}

We plot the the spatial profile $x_{i}(T)$  as a function of
site index $i$(See Fig. \ref{fig14}) for large $T$ in critical region. 
We simulate $N=100$ sites. The spatial profile is plotted
for $T=10^6$ and $T=10^6+1$
for $\epsilon=0.13,0.15,0.18$. We observe a zigzag pattern for 
$\epsilon=0.13$ and for $\epsilon=0.15$ but not for 
$\epsilon=0.18$.

\begin{figure*}[h]
       \includegraphics[scale=0.3]{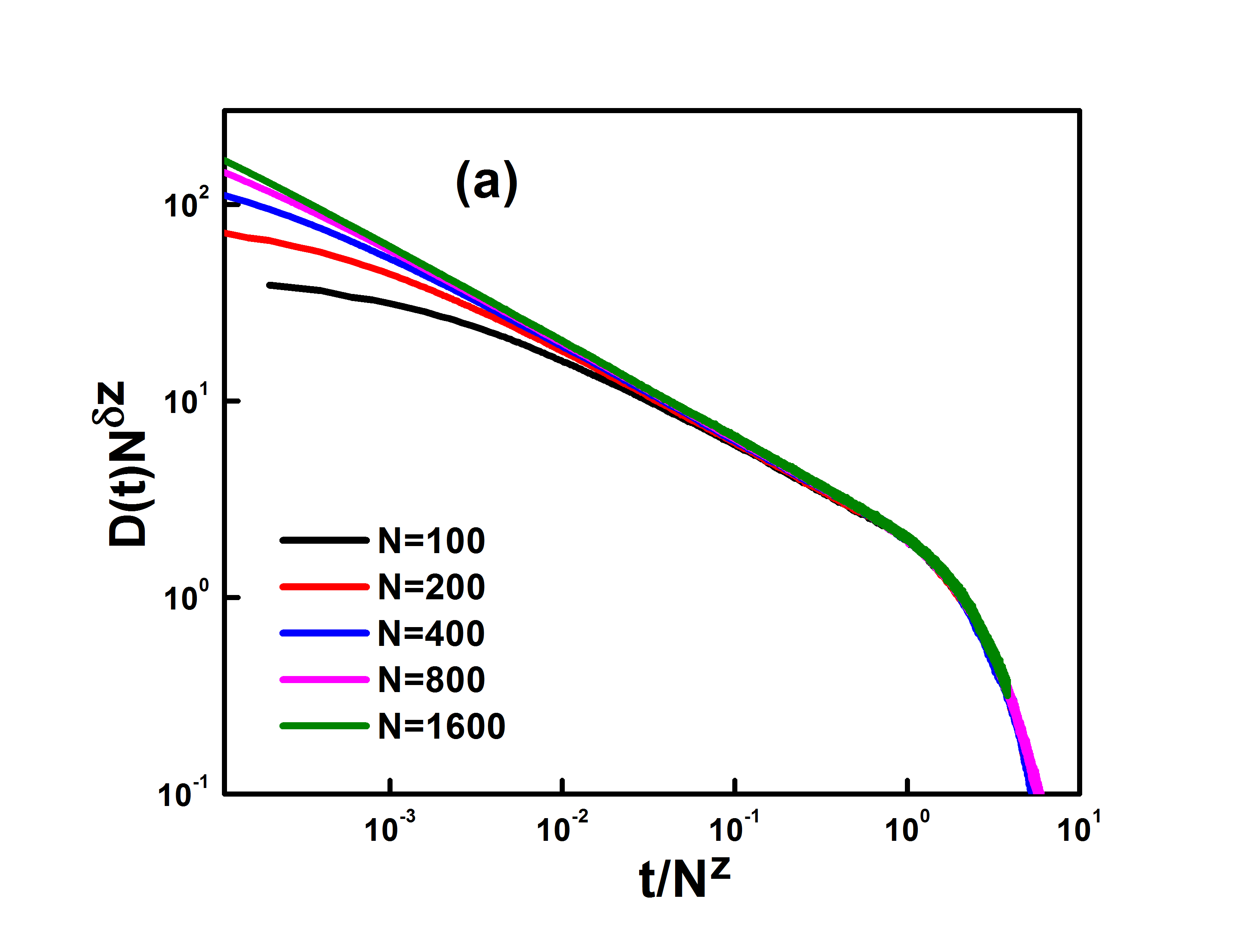}
       \includegraphics[scale=0.3]{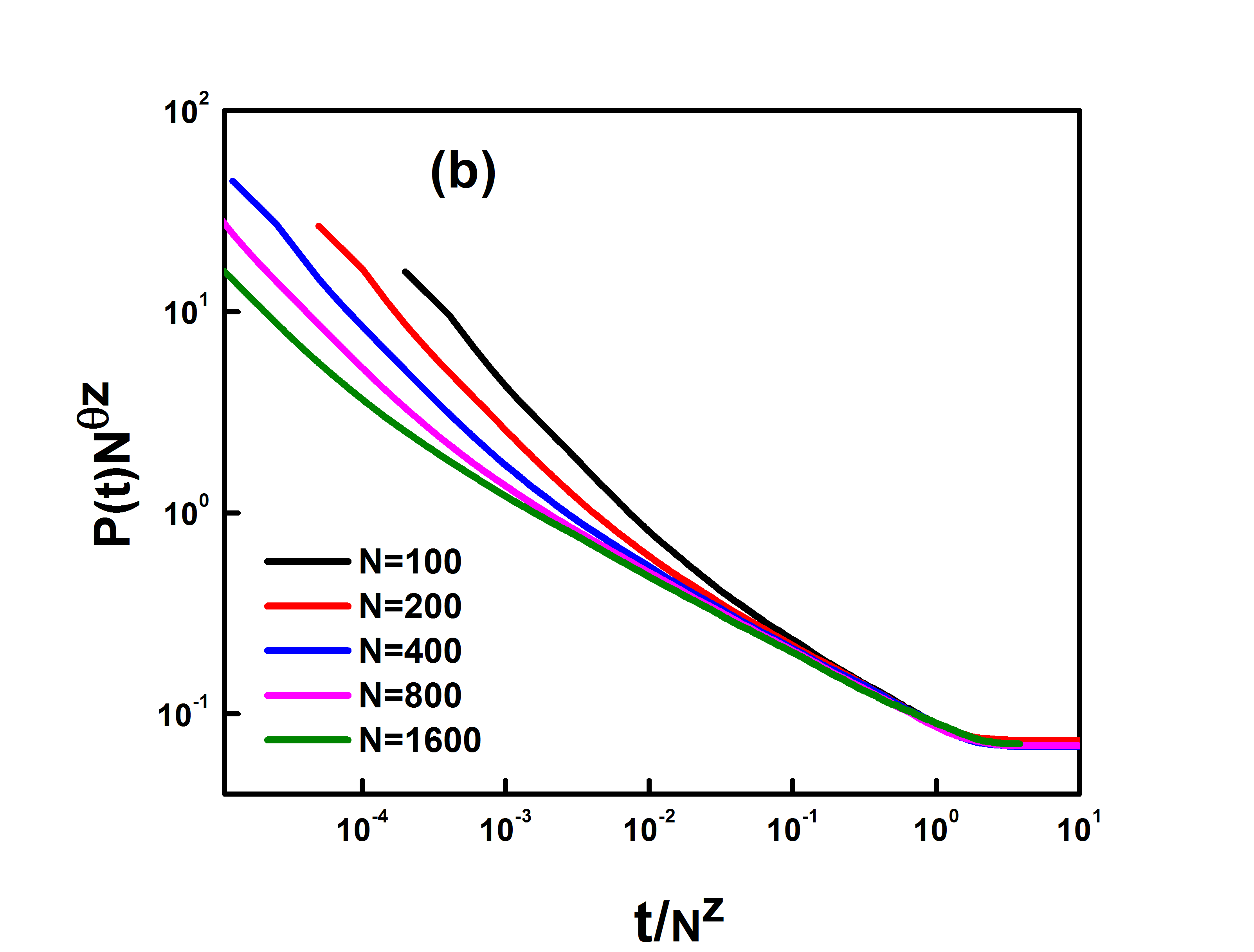}
	\caption{(a) We plot $D(t)N^{\theta z}$ for $N=2^m \times 100$,
	$m=0-4$ $N$ as a function of $t/N^z$ at $\epsilon=\epsilon_c=0.15$. 
	The excellent scaling collapse is obtained at $z=2$ and $\delta=0.5$.
	(b) We plot $P(t)N^{\theta z}$ for $N=2^m \times 100$, $m=0-4$ as 
	a function of $t/N^z$ at $\epsilon=\epsilon_c=0.15$.
	The fine collapse is obtained at  $z=2$ and $\theta=0.375$ }
       \label{fig15}
\end{figure*}

We carry out a finite-size scaling to compute the value of
dynamic exponent $z$. We consider $N=2^m \times 100$, $m=0-4$
and each average over  $5\times 10^3$ configurations.
We have plotted $D(t)N^{\delta z}$ as a function of
$t/N^z$ at $\epsilon=\epsilon_c=0.15$ (See Fig.\ref{fig15}(a)).
The fine collapse is obtained at $z=2$. In Fig.\ref{fig15}(b),
we plot $P(t)N^{\theta z}$ as a function of $t/N^z$ at
$\epsilon=\epsilon_c=0.15$. The good scaling collapse is obtained at $z=2$ in one dimension.

\begin{figure*}[h]
       \includegraphics[scale=0.3]{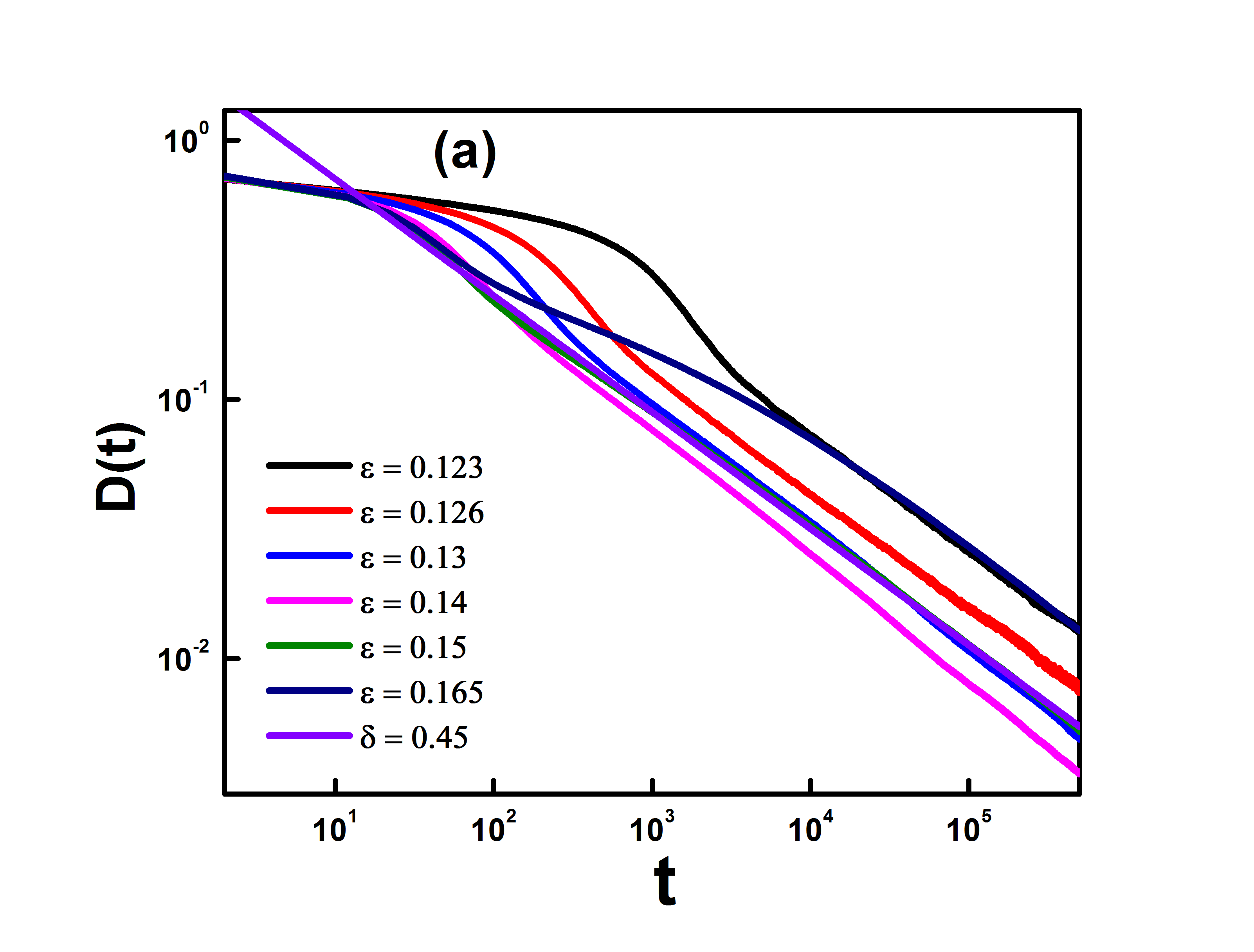}
       \includegraphics[scale=0.3]{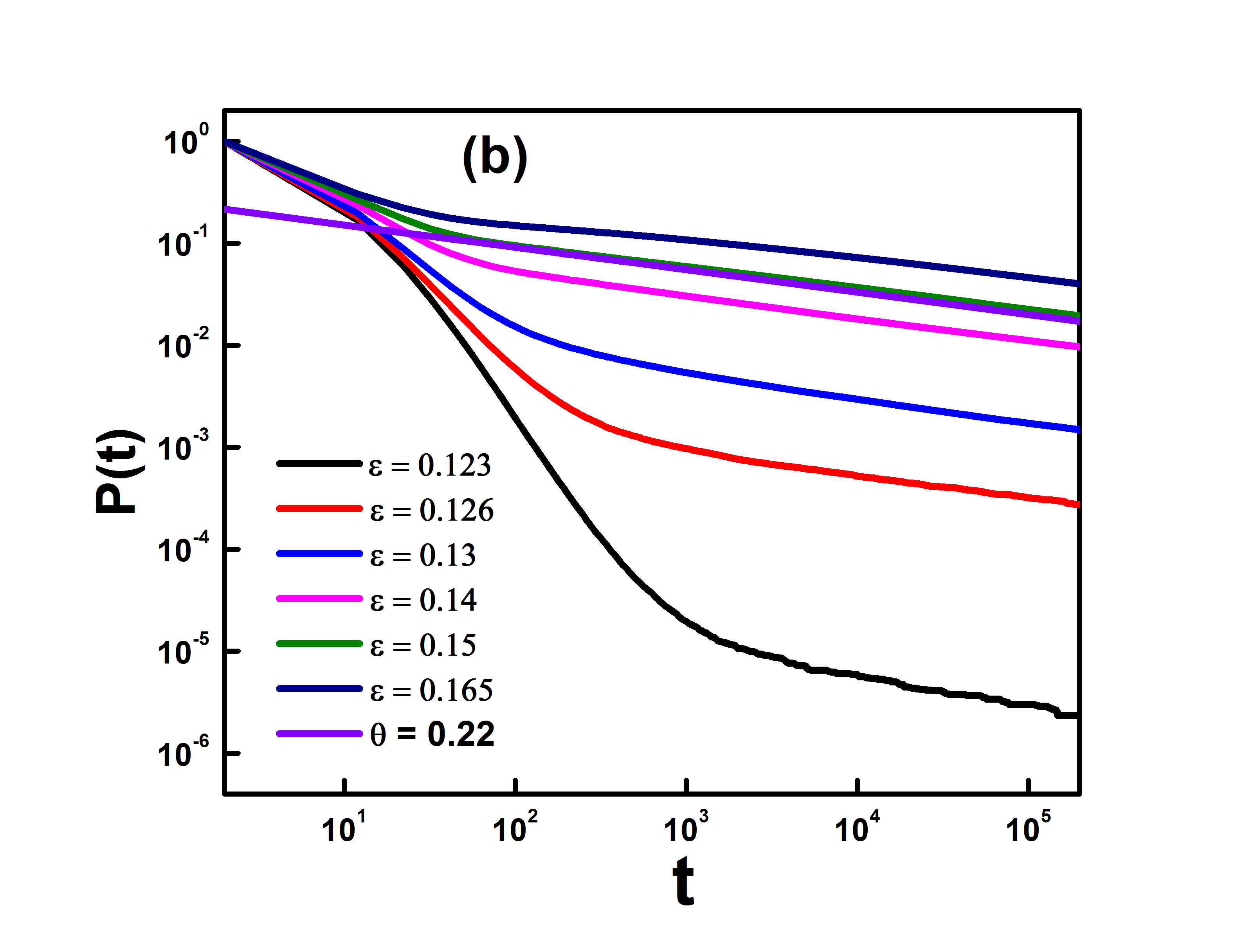}
       \caption{For coupled logistic maps in two dimensions,
	we plot phase defect $D(t)$ and phase persistence $P(t)$ as
	a function of time $t$. We consider $10^3\times 10^3$ lattice
	and averaged over 10  configurations for $\epsilon_1<\epsilon<\epsilon_2$ where $\epsilon_1=0.123$ and 
	$\epsilon_2=0.165$. a) $D(t)$ shows power-law
	decay with exponent $\delta=0.45$. (b) $P(t)$ shows power-law  decay with exponent $\theta=0.22$. }
       \label{fig16}
\end{figure*}

Now, we study coupled logistic maps in two dimensions.
We consider $ N \times N$ lattice with $N=10^3$. We
average over 10 configurations and simulate for
$10^5$ time-steps. Similar to the case
of the Gauss map in two dimensions, we observe power-law decay of $D(t)$ and
$P(t)$ over the range of coupling $\epsilon$ such that
$\epsilon_1 < \epsilon<\epsilon_2$ where $\epsilon_1=0.123$ to $\epsilon_2=0.165$.
The decay exponent of $D(t)$ is $0.45$ (See Fig.\ref{fig16}(a)) and for $P(t)$  is 0.22 (See Fig.\ref{fig16}(b)) which are
the same as for Gauss maps. These exponents do not change throughout the critical range.

\begin{figure*}[h]
       \includegraphics[scale=0.3]{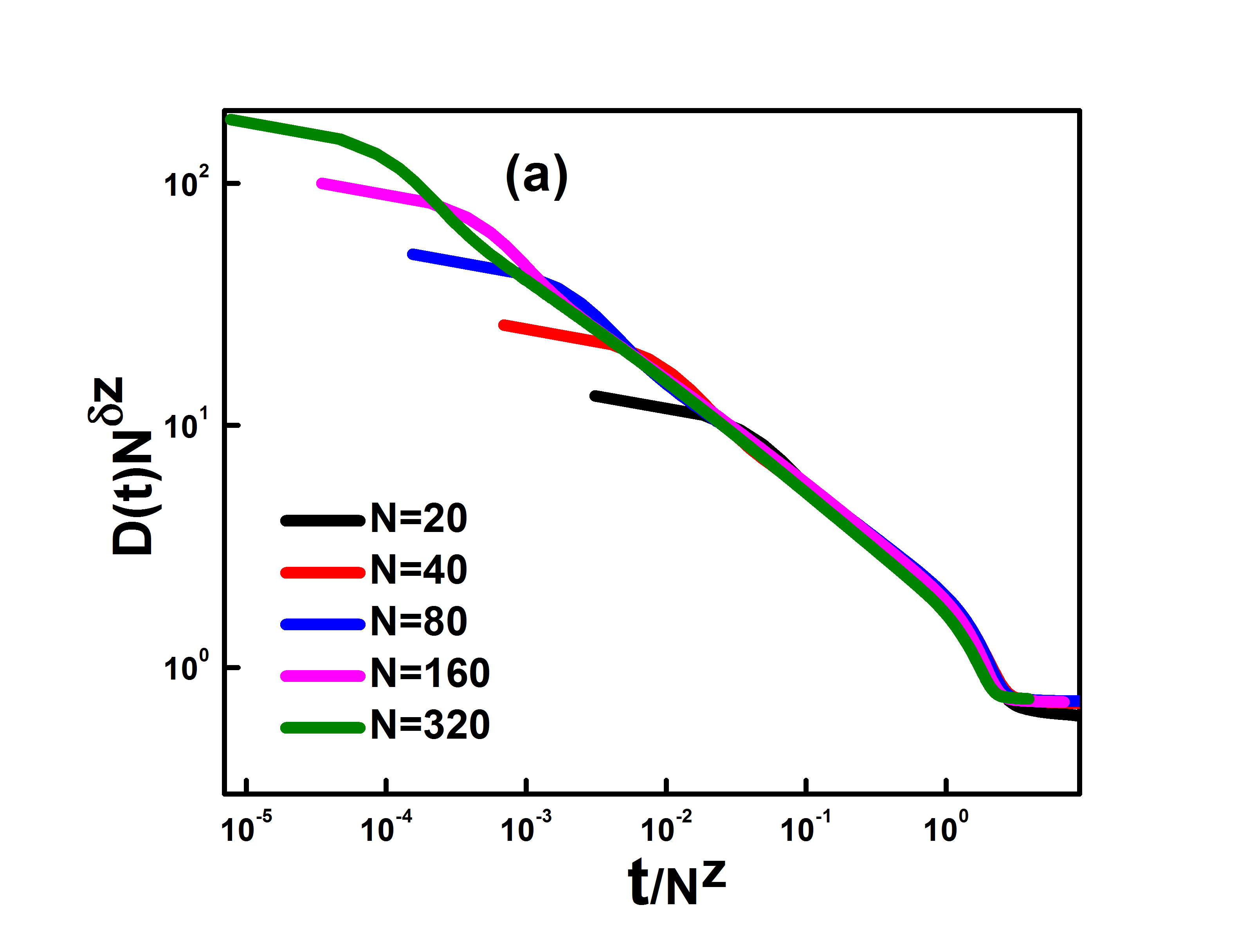}
       \includegraphics[scale=0.3]{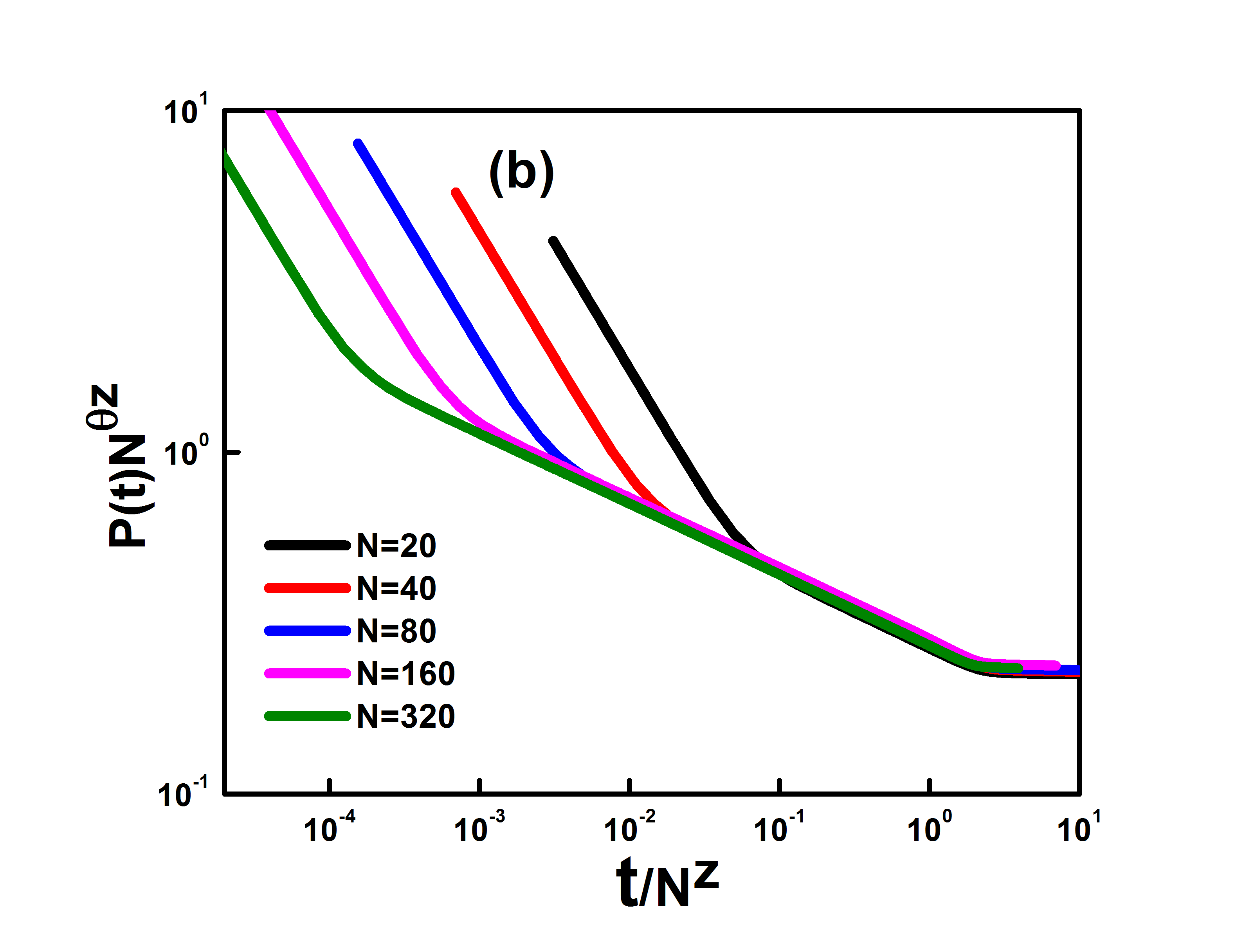}
       \caption{For $\epsilon=0.15$, we simulate coupled logistic maps 
	on $N\times N$ lattice for $N=2^m\times 10, m=1-5$. We each average over $4\times 10^5$ configurations.
	a) We plot $D(t)N^{\delta z}$ as a function of $t/N^z$.
	A fine collapse is obtained at $z=2.16$ and  $\delta=0.45$.
	b) We plot $P(t)N^{\theta z}$ as a function of $t/N^z$.
	A fine collapse is obtained at $z=2.16$ and $\theta=0.22$. }
       \label{fig17}
\end{figure*}

We carry out a finite-size investigation for a 2-d lattice. We simulate the
$N\times N$ lattice, with $N=2^m\times 10,m=1-5$ and each average over more than $4\times 10^5$ configurations. We have plotted $D(t)N^{\delta z}$ as a function
of $t/N^z$ (see Fig.\ref{fig17}(a)). The fine collapse is obtained at $z=2.16$ for $\delta=0.45$ and $\epsilon_c=0.15$.
Similarly, We also plot $P(t)N^{\delta z}$ as a function
of $t/N^z$ (see Fig.\ref{fig17}(b)) for the same size and configurations.
The fine collapse is obtained at $z=2.16$ for $\theta=0.22$
and $\epsilon_c=0.15$. 
Unlike the case of Gauss maps, we observe a clean collapse 
at the value expected for the Ising model. The reason could be
an absence of a long-lived metastable state at long times in this case.

The comparison with the Ising model may be in order.
A 2-d  Ising model with nonconserved order parameter will be an appropriate comparison. 
The dynamic exponent, in this case, is $z=2.16$ which matches the above model \cite{PhysRevE.101.012122}.
The above exponent has been obtained in Monte-Carlo simulation of a 2d Ising model using a heat-bath algorithm \cite{lei2007monte},  
finite-size scaling of 2d Ising model using
correlation times \cite{PhysRevLett.76.4548}
as well as using the damage spreading method \cite{FWang_1995}. 
A quench at a critical temperature from the initial random state leads to
correlation length growing with exponent $1/z=0.46$
\cite{walter2015introduction}. This exponent is close to the exponent with which we observe the decay of defects. The persistence exponent $0.22$  is also close to one obtained for the Ising model\cite{Jain_2000}. 
The glaring difference is obviously that these
exponents have been observed over a range of parameters and not a single point. Besides, these are two different points because the persistence exponent is expected only at zero temperature which is not
the critical temperature in two dimensions.

\begin{figure*}[h]
       \centering
       \includegraphics[scale=0.3]{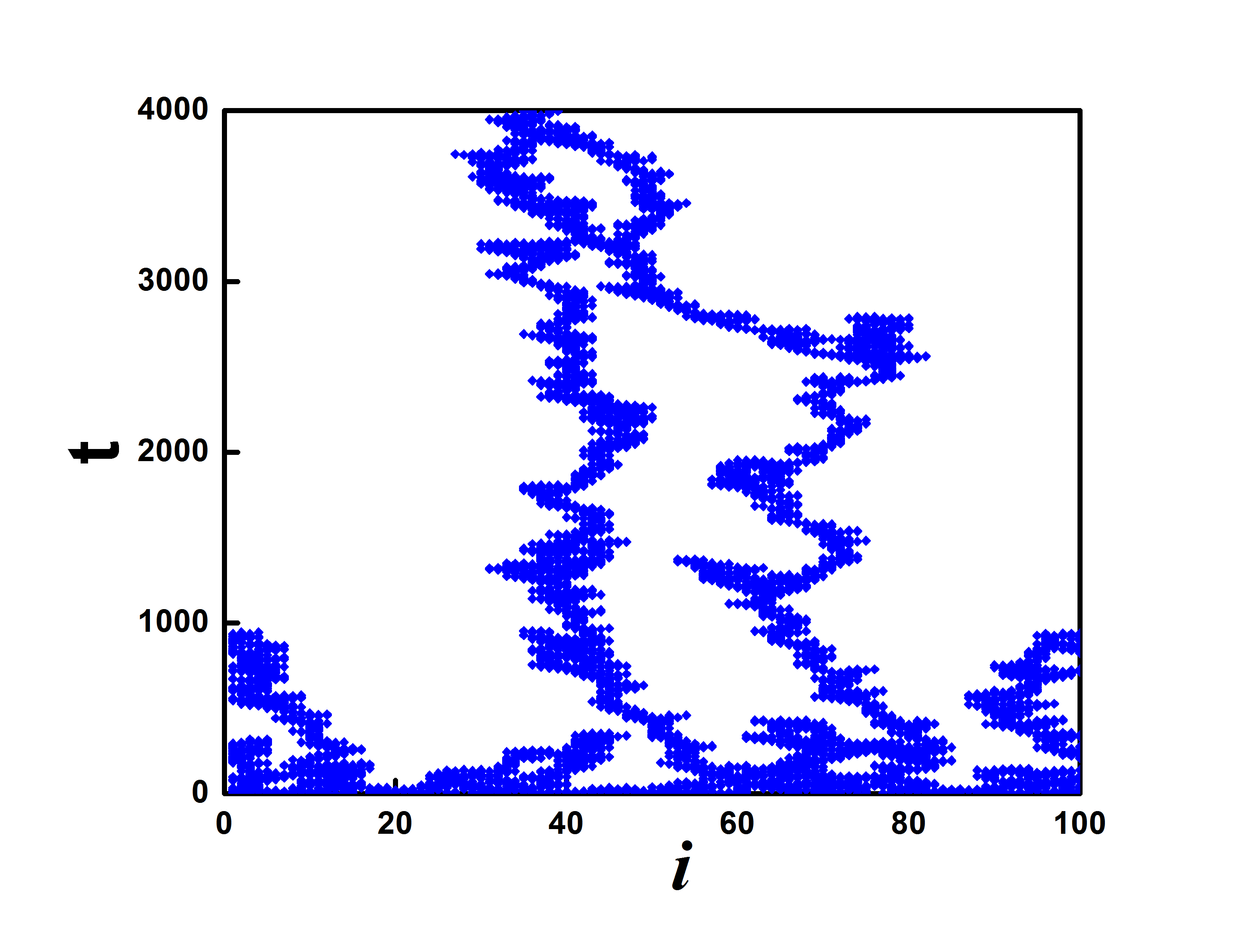}
       \caption{We plot spatiotemporal dynamics of defects as a
	function of time at $\epsilon$ = 0.13.}
       \label{fig18}
\end{figure*}

We carry out visualization of dynamics for coupled logistic maps
for $\epsilon=0.13$ in the critical region in a manner analogous to coupled Gauss maps.
In Fig.\ref{fig18} we plot the position of defects (consecutive sites with the same spin) in time.  We observe that defects carry out random walks, and annihilate each other to obtain
lattice with effective antiferromagnetic ordering.  

\begin{figure*}[h]
       \includegraphics[scale=0.195]{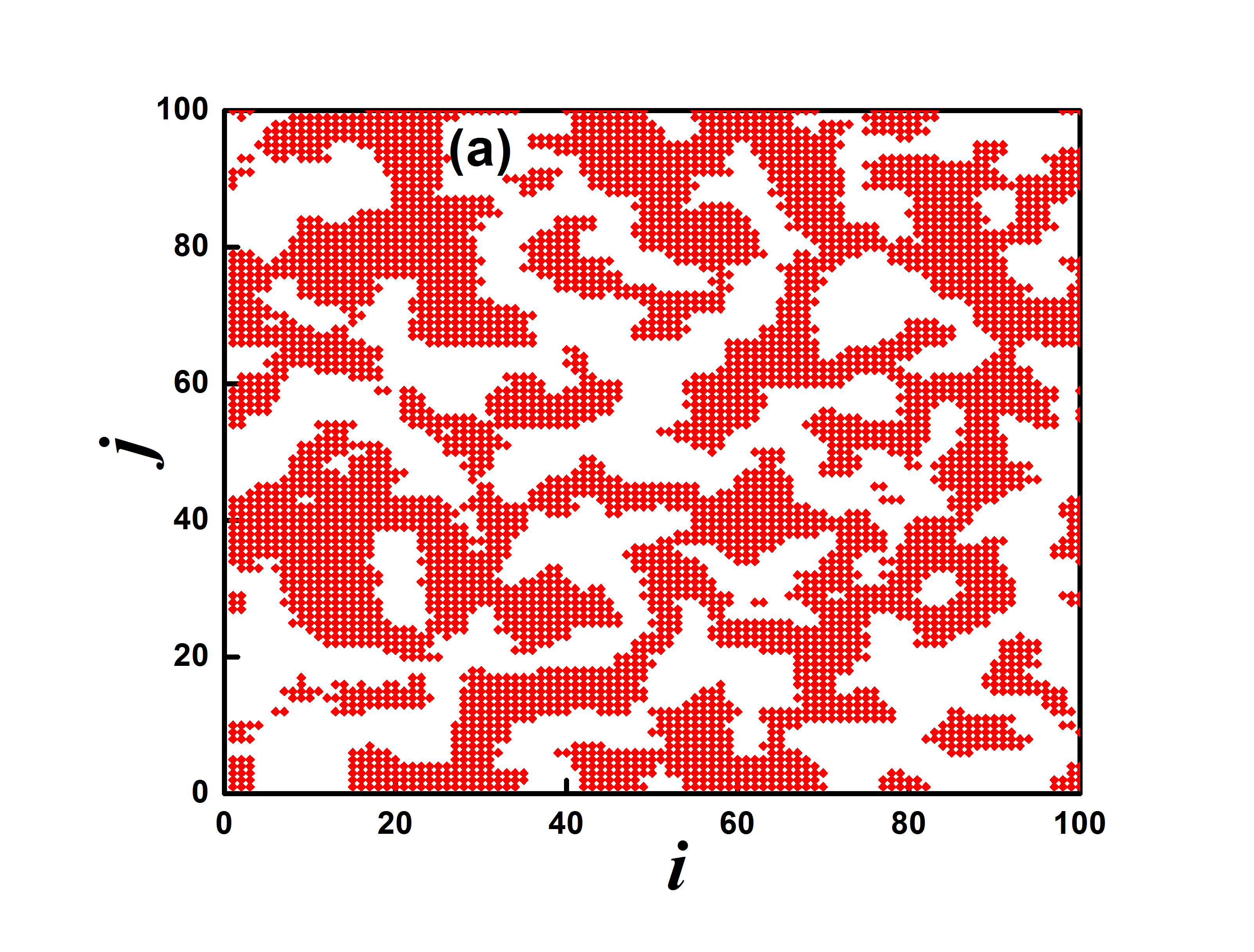}
       \includegraphics[scale=0.195]{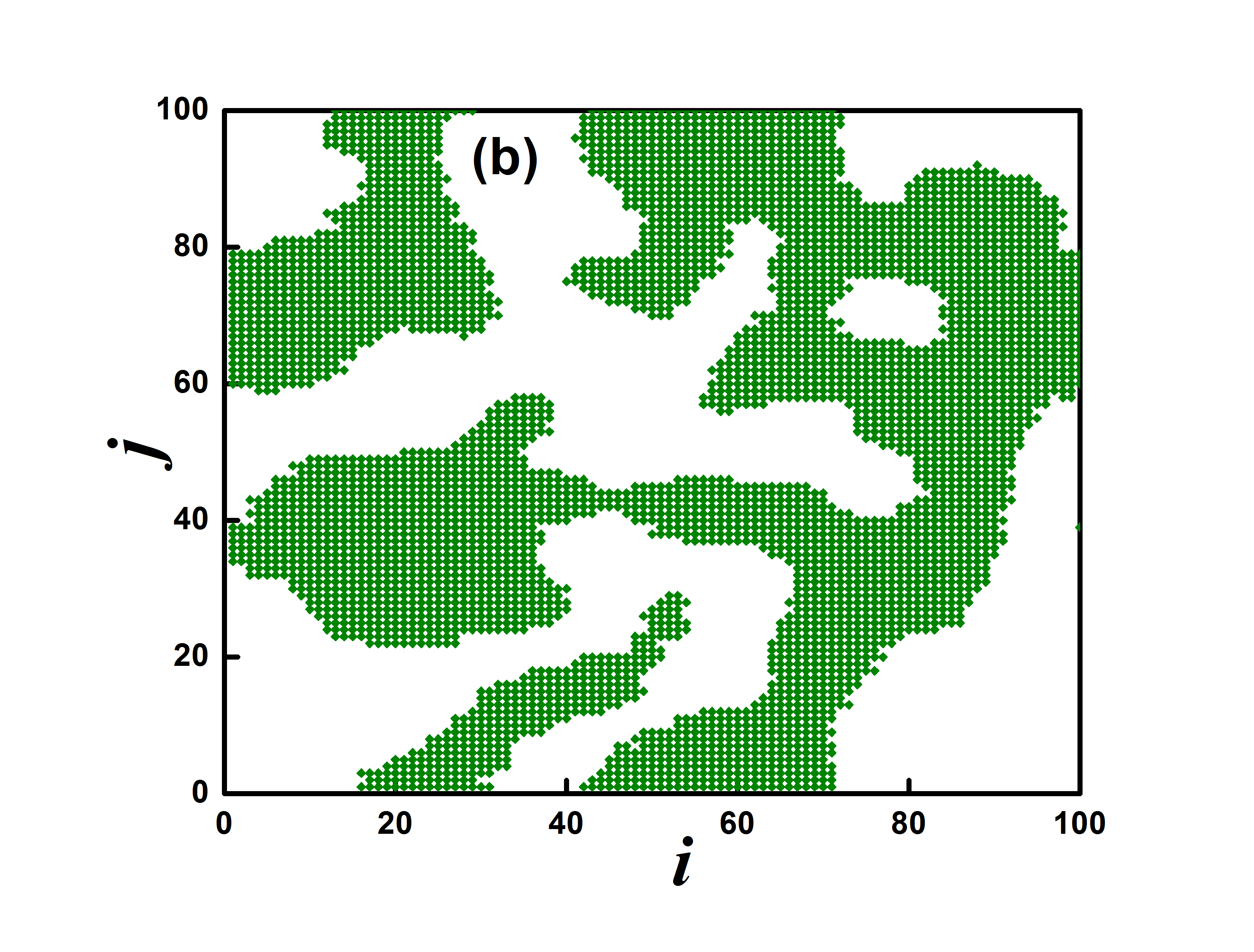}
       \includegraphics[scale=0.195]{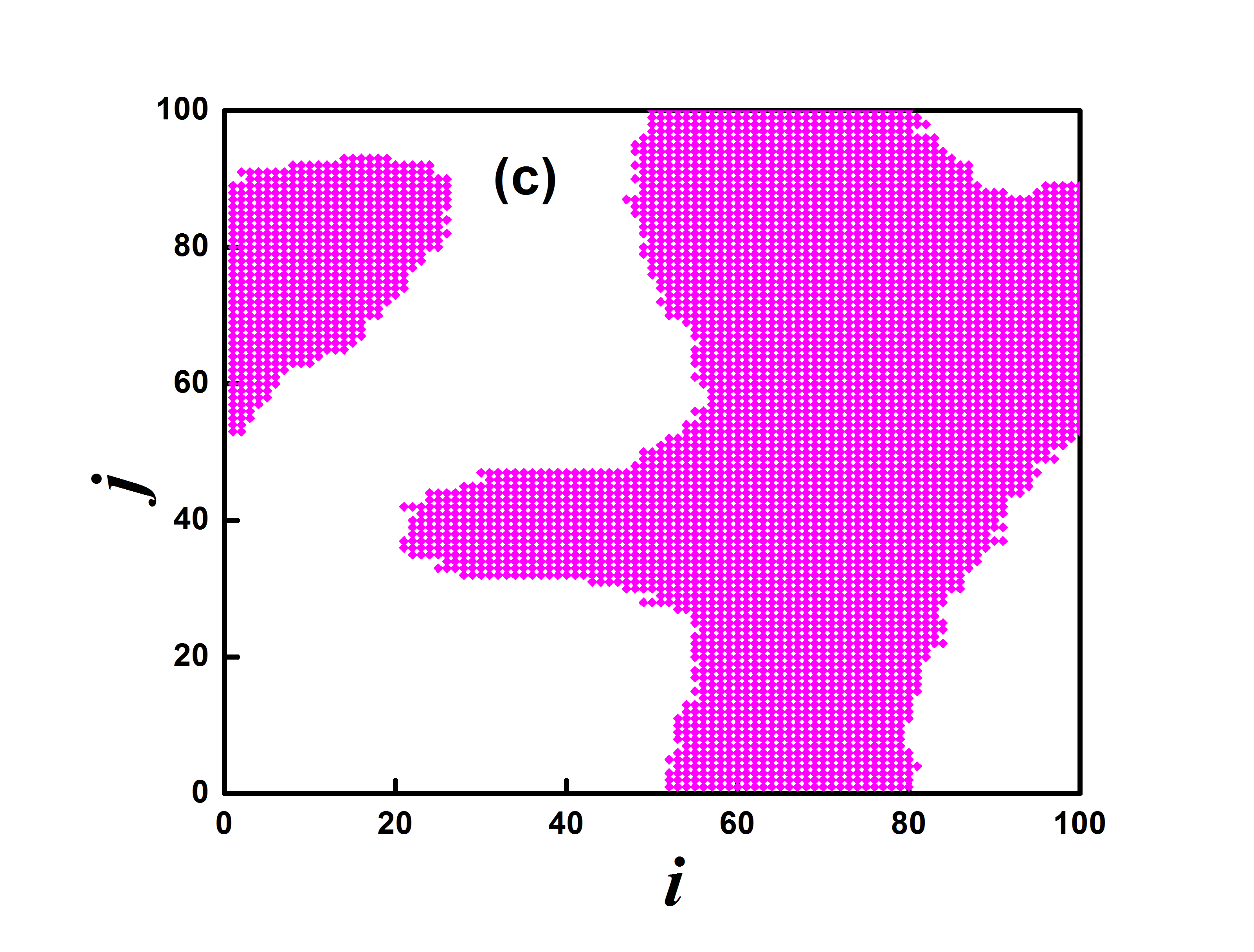}
       \caption{For logistic map, we plot sites $(i,j)$ such that
        $(-1)^{i+j} s(i,j)=1$ for
        a) $t$ = $10^2$ , b)  $t$= $10^3$, and
        c)  $t$= $10^4$ }
       \label{fig19}
\end{figure*}

To visualize the phase in two dimensions, we again
change the sign of the odd sublattice.
For $N=100$ and  $\epsilon=0.15$, we plot sites $(i,j)$ such that $(-1)^{i+j}s_{i,j}=1$.
The extent of the checkerboard pattern (ferromagnetic pattern in this representation)
grows in time and finally forms a giant phase. Three such representations are plotted in Fig.\ref{fig19}.

\section{Summary}
We study the zigzag pattern in one dimension or the checkerboard
pattern in two dimensions for coupled Gauss map and logistic
map. We introduce phase defect and phase persistence as order parameters to quantify the phase transition and associated universality class.

(a) In a one-dimensional coupled Gauss map and logistic map, 
we find the power-law decay of phase defect and phase persistence
over the range of critical parameters. In the Gauss map,  $\epsilon$ ranges from $\epsilon_1=-2.61$ to $\epsilon_2=-1.26$.
In the logistic map, it ranges from $\epsilon_1=0.124$  to $\epsilon_2=0.164$. In both maps, the decay exponent of $D(t)$ is found to be $\delta=0.5$ and for $P(t)$ the exponent is $\theta=0.375$. The dynamic exponent at the critical point is found to be $z=2$ in both the Gauss map and logistic map for both $P(t)$ and $D(t)$. 

(b) We extend our definition in two dimensions for coupled Gauss map
and logistic map. We find the power-law decay of phase defect and phase persistence over the range of coupling parameters.
In two dimensional coupled Gauss map, the range of $\epsilon$ is from $\epsilon_1=-2.1$ to $\epsilon_c=-1.25$. 
In two dimensional logistic map, it ranges from $\epsilon_c=0.123$ to $\epsilon_c=0.165$. In both maps, the decay exponent for $D(t)$ and $P(t)$ over the entire range is
found to be $\delta=0.45$ and $\theta=0.22$. For the Gauss map, we observe two
kinks and at the first kink, the obtained value of the dynamic exponent is 
$z=2$ for both $D(t)$ and $P(t)$. For the logistic map, the obtained value of 
the dynamic exponent is $z=2.16$ for both $D(t)$ and $P(t)$. 

Several dynamic phase transitions can be identified from bifurcation
diagram alone.
However, it is difficult to conclude the zigzag or checkerboard patterns noted
in this work from the bifurcation diagram alone. Thus these quantifiers which do
not require detailed knowledge of underlying equations are useful.
The observation of power law over a range of parameters and values similar
to those obtained for the Ising model are unexpected results in our work.

\section{Acknowledgment}
PMG thanks DST-SERB (CRG/2020/003993) for financial assistance.
MCW thanks the Council of Scientific and Industrial Research (C.S.I.R.),
SRF (09/128(0097)/2019-EMR-I).









\section{}

\printcredits

\bibliographystyle{unsrt}

\bibliography{ref}

\bio{}
\endbio

\bio{}
\endbio

\end{document}